\documentclass[12pt, draftclsnofoot, onecolumn]{IEEEtran}

\usepackage{graphicx}

\usepackage{subfigure}

\usepackage{amsmath,amssymb}\usepackage{amsthm}
\usepackage[mathscr]{eucal}
\usepackage{graphics,graphicx,multicol}
\usepackage{color}
\usepackage{amsfonts}
\usepackage{epsfig}
\usepackage{cite,xspace,syntonly,algorithm,algorithmic}
\usepackage{bm}
\usepackage{setspace}
\newcommand{\Reals}     {{{\mathrm I\!R}}}  

\newcommand{\define}    {\stackrel{\scriptscriptstyle\triangle}{=}}  

\newcommand{\uwti}[1]{{\mathbf #1}}

\newcommand{\eb}{{\uwti e}}

          \newcommand{\Gammab}   {{\bm \Gamma}}

\newcommand{\Ac} {{\mathcal A}}         
\newcommand{\Bc} {{\mathcal B}}

\newcommand{\Ec} {{\mathcal E}}         
\newcommand{\Fc} {{\mathcal F}}

\newcommand{\Mc} {{\mathcal M}}

\newcommand{\Pc} {{\mathcal P}}

\newcommand{\Sc} {{\mathcal S}}         
         
\newcommand{\Uc} {{\mathcal U}}

\newcommand{\Iulk} {{\underline{{\bm {\mathcal I}}}}}

\newcommand{\Julk} {{\underline{{\bm {\mathcal J}}}}}

\newcommand{\Oulc} {{\underline{\mathcal O}}}
\newcommand{\Aulc} {{\underline{\mathcal A}}}

\newcommand{\Iulc} {{\underline{\mathcal I}}}
\newcommand{\Julc} {{\underline{\mathcal J}}}
\newcommand{\Gulc} {{\underline{\mathcal G}}}

\newcommand{\Eulc} {{\underline{\mathcal E}}}
\newcommand{\Fulc} {{\underline{\mathcal F}}}

\newcommand{\Sulc} {{\underline{\mathcal S}}}

              \newcommand{\hul}  {{\underline h}}
\newcommand{\Iul}  {{\underline I}}              
\newcommand{\Jul}  {{\underline J}}

\newcommand{\Omegaul}  {{\underline \Omega}}
\newcommand{\ba}{\begin{array}}
\newcommand{\ea}{\end{array}}

 \newtheorem{definition}{Definition}
\newtheorem{corollary}{Corollary}

\newtheorem{theorem}{Theorem}
\newtheorem{lemma}{Lemma}
\newtheorem{propo}{Proposition}

\begin{document}

\title{Exploiting Dual Connectivity in Heterogeneous Cellular Networks} 
\author{\IEEEauthorblockN{Narayan Prasad\IEEEauthorrefmark{1},
Sampath Rangarajan\IEEEauthorrefmark{2}}\\
\IEEEauthorblockA{\IEEEauthorrefmark{1}Futurewei, Bridgewater, NJ; \IEEEauthorrefmark{2}NEC Labs America, Princeton, NJ}\\
e-mail:   \{narayan.prasad1@huawei.com, sampath@nec-labs.com\}\thanks{This work was done when Narayan Prasad was with NEC Labs America.}  
} 

\maketitle
\thispagestyle{empty}
\begin{abstract}
We consider network utility maximization problems over heterogeneous cellular networks (HetNets) that permit dual connectivity. Dual connectivity (DC) is a feature that targets emerging practical HetNet deployments that will comprise of  non-ideal (higher latency) connections between transmission nodes, and has been recently introduced to the LTE-Advanced standard. DC allows for a user to be simultaneously served by a macro node as well as one other (typically micro or pico) node  and requires relatively coarser level coordination among serving nodes. For such a DC enabled HetNet we
 comprehensively analyze  the problem of determining   an optimal user association, where in any feasible association each user can be associated with (i.e., configured to receive data from) any one macro node (in a given set of macro nodes) and any one pico node that lies in  the chosen macro node's coverage area. We consider 
  the weighted sum rate system utility  subject to per-user maximum and minimum rate constraints, as well as the  proportional fairness (PF) system utility. For both utility choices we construct approximation algorithms and establish their respective approximation guarantees.  
 We then validate the  performance of our algorithms via numerical results. 
\end{abstract}
\pagestyle{empty}

\section{Introduction}
Traditional cellular wireless are rapidly transforming into dense HetNets that have discarded  the  classical   structured layout of cells. Instead, these HetNets are characterized by the presence of a multitude of transmission nodes (or points) ranging from enhanced versions of the conventional high power macro base-station or NodeB (eNBs) to low power pico nodes, all  deployed in a highly irregular fashion \cite{andrews:LBmag}. Indeed, the deployment of the low power nodes is done within the coverage area of an eNB  to cater to emerging hot spots, thereby alleviating demand bottlenecks without being subject to many of the challenges in eNB site acquisition.  However, a major hinderance in such deployments  is that there is need for coordination among the transmission nodes (which becomes more acute as the density of such nodes rises) while at the same time the
backhaul link between these nodes is often non-ideal. Consequently,
for tractable resource allocation, a HetNet is partitioned into several coordination units or clusters with each cluster comprising of a set of high power eNBs  along with a set of low power pico nodes assigned to each one of the high power nodes. Together, these transmission points (TPs) in each cluster cater to a given set of users. 
In addition, only semi-static coordination among TPs in a cluster is deemed feasible, wherein periodically (once in every frame of a few hundred milliseconds duration) there is coordination among serving nodes in the cluster. One popular method of coordination is load balancing or user association \cite{andrews:LBmag}
  where each user can be associated with only one TP at   any given time.
 This load balancing requires limited coordination among TPs  which is possible under a non-ideal backhaul, and it mitigates the undesirable scenario  of  TPs  becoming overloaded due to too many users being associated with them. Combinations of load balancing with several resource management schemes have also received wide attention \cite{altman:sinr,shen:lb,prasad:lb,bedekar:wiopt,ye:onoff,sonvec:ualb}. 


Our interest in this work is on dual connectivity (DC) that has been recently introduced to the 3GPP LTE-A standard \cite{3gpp}, where the single-TP association constraint is relaxed and a user can be associated to a high power and a low power node. Such a user can simultaneously receive (different) data from both nodes. Schemes to fully exploit  DC are being actively investigated and the potential challenges and good directions are summarized in \cite{jha:dc}. The work in \cite{wu:dc,wu:Icc} considers a DC enabled uplink with one macro and one pico node and proposes optimal rate and power control solutions for  a cost minimization problem with  per-user minimum rate constraints. Power optimization over a DC enabled Hetnet  has been considered in \cite{Sahmad:DC} where distributed algorithms for the uplink that account for backhaul capacity have been proposed. Power optimization is also investigated in \cite{Wu:nomadc} where non-orthogonal multiple access (involving successive interference cancellation at the receiving nodes) was additionally exploited in a DC enabled downlink comprising of a single macro base station and a single small cell access point
  to improve the throughputs. On the other hand, \cite{singh:dc} considers resource partitioning  at only the macro node in a DC enabled downlink to optimize the PF utility. \cite{kim:DCUA} proposes an efficient sub-optimal algorithm for the problem of determining the sum rate maximizing user association under the restrictions that for each user only the low-power node yielding the highest SINR can be chosen and each node employs round robin scheduling.  
\cite{Legg:DC} evaluates algorithms that aim to maximize the number of satisfied users, .i.e., aim to satisfy minimum rate requirements for as many users as possible via single-TP association as well as DC. \cite{Legg:DC} shows that a smartly designed heuristic to exploit DC can significantly improve user satisfaction. We note that our work  formally captures the notion of considering the association of all users to optimize a system utility (which also incorporates user satisfaction). \cite{lema:dc} employs stochastic geometry based tools to demonstrate the benefits of dual connectivity together with decoupled associations in the uplink and downlink.   
\cite{wang:dc} reuses existing algorithms for user association and investigates data forwarding and flow control problems, whereas packet scheduling algorithms for expoiting DC in the downlink have been proposed in \cite{pan:ddc}.  
 In this work we consider a general DC enabled HetNet downlink with multiple users and TPs. Our key contributions are  the following: 
\begin{itemize}
\item We propose an efficient algorithm that yields a user association that is optimal for the PF system utility up-to an additive constant. To the best of our knowledge, this is the first such approximation algorithm for  DC and  the PF utility. Using this algorithm, we demonstrate the significant gains enabled by DC especially at low network loads.  
\item We also show that the user association problem to optimize the weighted sum rate utility subject to per-user minimum and maximum rate constraints can be formulated as a {\em constrained non-monotone submodular set function maximization}.  This allows us to derive an efficient algorithm which guarantees a constant-factor approximation. We note that to show submodularity we  do not  follow the direct approach of proving the original definition, but instead  we consider  proving another cleverly obtained sufficient condition. The latter approach  then  requires us to characterize the (second order) change in the optimal solution of a linear program with respect to its parameters. We expect that our result and systematic derivation will have much wider applications.
\end{itemize}
 We note that a conference version of this report appeared in \cite{prasad:dc}. Compared to \cite{prasad:dc}, here we enhance the technical results, provide complete proofs as well as expand on simulation results,  discussions and new applications. 
\section{Problem Formulation}\label{sec:probform}
\begin{figure}[htbp]
    \centering
   \includegraphics[clip, trim=5cm 5cm 5cm 5cm,width=0.6\textwidth]{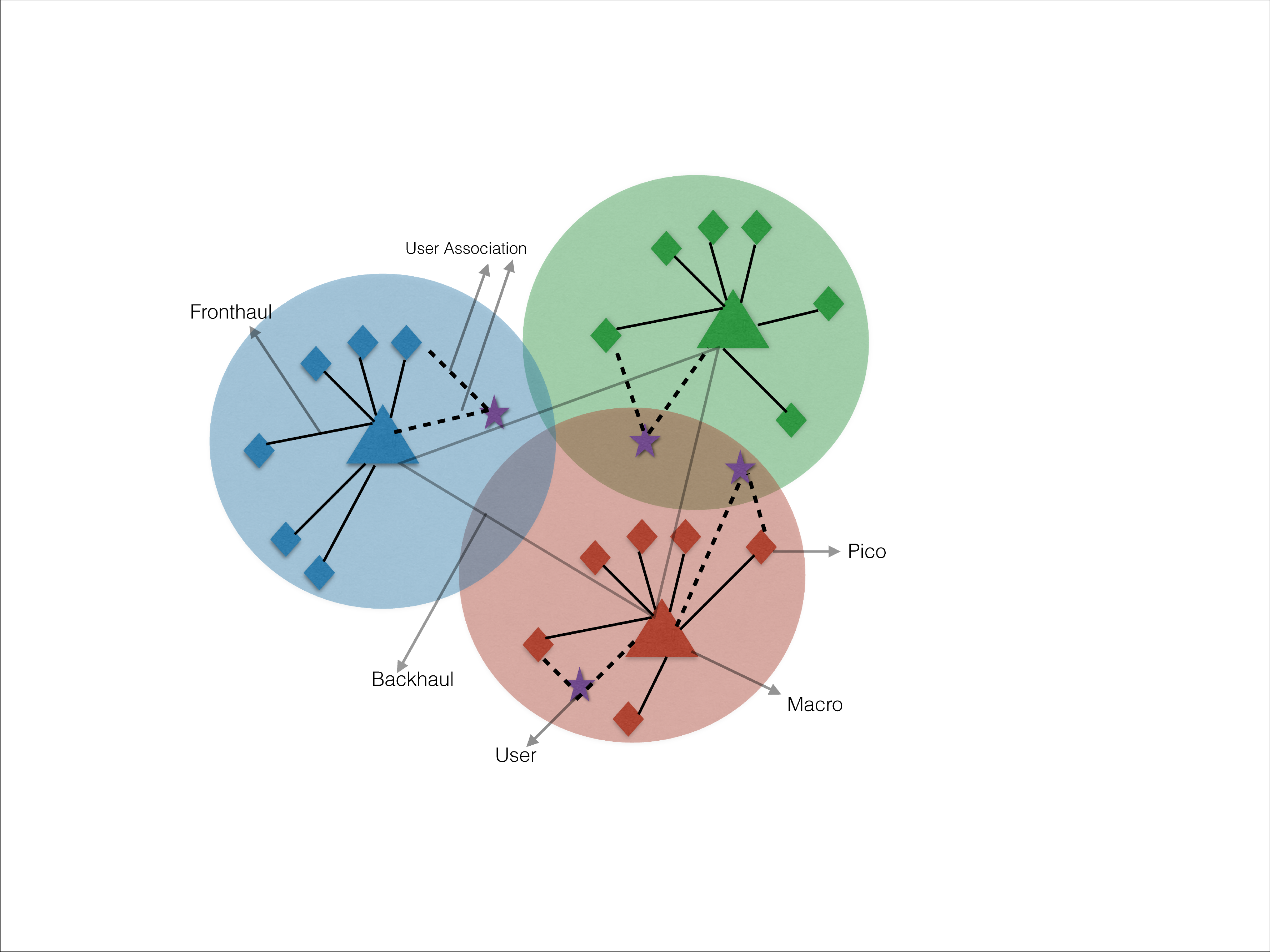}
\caption{Dual Connectivity Schematic}\label{fig:DCscheme} \vspace{-.5cm}
\end{figure}

Let $\Uc$ denote the set of users with cardinality $|\Uc|=K$ and let  $\Mc$ denote the set of Macro TPs. For each Macro
 TP $m\in\Mc$, let $\Bc_m$ denote the set of pico TPs assigned to macro TP $m$. Here the set $\Bc_m$ of pico TPs facilitate the macro TP $m$ to serve its associated users.  We suppose that all indices in 
  the set of all TPs, $\Sc = \Mc\cup \Pc$, where $\Pc=\cup_{m\in\Mc}\Bc_m$, are distinct.
 An illustrative schematic for DC is shown in Fig. \ref{fig:DCscheme}.
  Notice that each user can be associated with any one macro and any one pico TP from the set of pico TPs assigned to the chosen macro.
For each user $u\in\Uc$ and each  TP $b\in\Sc$, we define $R_{u,b}$ to be the averaged single-user or peak rate user $u$ can get (in bits per unit resource) when it alone is served by TP $b$. This single-user or peak rate is a function of the slow fading parameters (e.g. path loss and shadowing) seen by user $u$ but not the instantaneous fast-fading ones which have been averaged out. Further, this average rate can be computed under the (widely used) assumption that all the other interfering  TPs    are always transmitting, or under the framework of \cite{singh:af} that each interfering TP transmits for a given fraction of the frame  duration. 
We note that in the in-band DC case the picos and macros share the same spectrum band whereas in the out-of-band DC case they are assigned different bands. Our  results are applicable to both scenarios and  only the user peak rates have to be accordingly computed.    
  We make the mild assumption that 
  $R_{u,b}>0,\;\forall\;u\in\Uc,b\in\Sc$. We also assume  that $\frac{R_{u,m}}{R_{u,b}}\neq \frac{R_{u',m}}{R_{u',b}},\;\forall\;b\in\Bc_m,m\in\Mc\;\&\;u\neq u'$. Notice that these two assumptions hold true almost surely for all typical slow fading distributions.
\begin{figure*}{{\begin{equation}\boxed{\begin{aligned}\label{eq:Bigoriginalwsr2}
   \max_{z_{u,m},x_{u,b}\in\{0,1\}, \gamma_{u,b},\theta_{u,m}\in[0,1]\;\forall\;u\in\Uc,b\in\Bc_m,m\in\Mc}\left\{\sum_{u\in\Uc}\sum_{m\in\Mc}\left(w_u R_{u,m}\theta_{u,m}+ \sum_{b\in\Bc_m}w_ux_{u,b}R_{u,b}\gamma_{u,b}\right)\right\}\\
 {\rm s.t. } \sum_{m\in\Mc}z_{u,m}\leq 1; \; \sum_{b\in\Bc_m}x_{u,b} = z_{u,m},\;\forall\;u\in\Uc,m\in\Mc;\\
 \sum_{u\in\Uc}\theta_{u,m}\leq 1\;\&\; \sum_{u\in\Uc}\gamma_{u,b}\leq 1\;\forall\;b\in\Bc_m,m\in\Mc;\\
   R_{u,m}\theta_{u,m}+ \sum_{b\in\Bc_m}x_{u,b}R_{u,b}\gamma_{u,b}\in [z_{u,m}R^{\rm min}_u, z_{u,m}R^{\rm max}_u],\;\forall\;u\in\Uc. \vspace{-.8cm}
\end{aligned}}\end{equation}}  \vspace{-1.3cm}}      \end{figure*}
We first formulate a network utility maximization problem that adopts the weighted sum rate (WSR)  system utility under per-user rate constraints.  
 The   WSR problem is posed in (\ref{eq:Bigoriginalwsr2}).
 $w_u>0,\;\forall\;u\in\Uc$ denotes any input weight or priority assigned to user $u$.
In (\ref{eq:Bigoriginalwsr2}) the binary valued variable $z_{u,m}$ is one if user $u$ is associated with macro TP $m$ and zero otherwise, so that the first set of constraints in (\ref{eq:Bigoriginalwsr2}) ensures that each user is associated with at-most  one macro. Further, exploiting dual connectivity,  each user that is associated to the macro TP $m$ is also associated with any one pico TP in $\Bc_m$. Indeed,
 the indicator variable
 $x_{u,b}$ is one if user $u$ is associated to TP $b$ and zero otherwise.  Consequently,  $\{x_{u,b}=1\}_{u\in\Uc,b\in\Bc_m}$ yields the user set such that each user in that set is associated to any one TP in $\Bc_m$ as well with the macro TP $m$. The continuous variables $\{\gamma_{u,b},\theta_{u,m}\}$ are referred to here as allocation fractions and their respective sums are upper bounded by unity for each macro TP as well as each pico TP, as depicted in the second set of constraints.  
 Although (\ref{eq:Bigoriginalwsr2}) does not enforce that each user must be associated with a macro TP, it does enforce (in the third set of constraints) that each user  associated to any macro TP   must be assigned at-least its minimum rate but should not exceed a maximum rate.  Notice also that (\ref{eq:Bigoriginalwsr2}) is always feasible. \footnote{Note that single-TP only association is subsumed by (\ref{eq:Bigoriginalwsr2}). Indeed, with $x_{u,b}=1$ for some $u\in\Uc,b\in\Bc_m$, by choosing $\theta_{u,m}=0$  ($\gamma_{u,b}=0$) we can ensure that user $u$ will receive data only from the pico node $b$ (macro node $m$). Also, each user $u$ that is not associated with any macro TP gets zero rate and must have $\theta_{u,m}=0,\gamma_{u,b}=0,\forall\;b\in\Bc_m,m\in\Mc$ due to the third set of constraints. } The minimum rate constraints are useful in enforcing that a minimum expected quality is provided to  each served user and are based on the observation that  in many scenarios serving a user at a rate below its minimum threshold is futile. 
 On the other hand,  the maximum rate limits can be used to cap the rates of any set of users such as those that have subscribed to a lower tier of service.
We next consider the PF system utility and adopting the convention that $0\ln(0)=0$, we pose
a  mixed optimization problem given in (\ref{eq:Bigoriginal}).
  \begin{figure*}{{\begin{equation}\boxed{\begin{aligned}\label{eq:Bigoriginal}
   \max_{z_{u,m},x_{u,b}\in\{0,1\}, \gamma_{u,b},\theta_{u,m}\in[0,1]\;\forall\;u\in\Uc,b\in\Bc_m,m\in\Mc}\left\{\sum_{u\in\Uc}\sum_{m\in\Mc}\sum_{b\in\Bc_m}x_{u,b}\ln\left( R_{u,m}\theta_{u,m}+ R_{u,b}\gamma_{u,b}\right)\right\}\\
 {\rm s.t. } \sum_{m\in\Mc}z_{u,m}= 1;  \sum_{b\in\Bc_m}x_{u,b}= z_{u,m},\;\forall\;u\in\Uc,m\in\Mc;\\
 \sum_{u\in\Uc}\theta_{u,m}\leq 1\;\&\; \sum_{u\in\Uc}\gamma_{u,b}\leq 1\;\forall\;b\in\Bc_m,m\in\Mc.
\end{aligned}}\end{equation}} \vspace{-1cm}}      \end{figure*}
The first set of constraints in (\ref{eq:Bigoriginal}) ensures that each user is associated with exactly  one macro. As before, exploiting DC each user that is associated with the macro TP $m$ is also associated with any one pico TP in $\Bc_m$.

Note that our formulations assume an infinitely backlogged traffic model with no limits on buffer sizes at any TP. Coordination among TPs happens at   frame boundaries where the user  association can be altered. After a transient phase (whose length can be ignored), for each user distinct data streams are available for downlink transmission at its assigned macro as well as its assigned pico node. This setting bestows tractability while being relevant. Extending our results to a more realistic formulation with finite buffers entailing  careful data forwarding (from each macro to each pico assigned to it)  is an interesting topic for future work. 
\section{Characterizing Optimal Allocation Fractions}\label{sec:copt}
We begin our quest by characterizing the optimal allocation fractions of (\ref{eq:Bigoriginalwsr2}) and (\ref{eq:Bigoriginal}) for any given  user association. This then enables the design of approximation algorithms. 

\subsection{Optimal allocation fractions of (\ref{eq:Bigoriginalwsr2})}\label{sec:charwsr}
We now proceed to characterize the solution of (\ref{eq:Bigoriginalwsr2}) for any given user association.
We note that upon fixing the user association in (\ref{eq:Bigoriginalwsr2}) (i.e., upon fixing $\{x_{u,b},z_{u,m}\}$) the problem in (\ref{eq:Bigoriginalwsr2}) decouples into
 $|\Mc|$ sub-problems, one for each macro TP. Consequently, we focus our attention  on the subproblem corresponding to any macro TP, say with index $1$,  and suppose
  that  any subset of users $\Uc'\subseteq\Uc$ is associated to that macro by the given association. Then, for each $b\in\Bc_1$, let $\Uc^{(b)}=\{u\in\Uc:x_{u,b}=1\}$ denote the associated user set such that
 $\Uc^{(b)}\cap\Uc^{(b')}=\phi,b\neq b'$, where $\phi$ denotes the empty set and $\cup_{b\in\Bc_1}\Uc^{(b)}=\Uc'$. Let $\Bc_1'$ denote the set of all pico TPs in $\Bc_1$ with at-least one associated user.
 In addition, we consider a  budget constraint for each pico TP, $\Gamma_b\in [0,1],\;\forall\;b\in\Bc_1'$ and another one for the macro, $\Gamma\in[0,1]$.
With these in hand we pose the problem in (\ref{eq:Bigoriginalwsrp}),
 \begin{figure*}{{\begin{equation}\boxed{\begin{aligned}\label{eq:Bigoriginalwsrp}
   \max_{\gamma_{u,b},\theta_{u,1}\in[0,1]\;\forall\;u\in\Uc',b\in\Bc_1'}\left\{\sum_{b\in\Bc_1'}\sum_{u\in\Uc^{(b)}}\left(w_u R_{u,1}\theta_{u,1}+ w_uR_{u,b}\gamma_{u,b}\right)\right\}\\
 {\rm s.t. }  \sum_{u\in\Uc'}\theta_{u,1}\leq \Gamma\;\&\; \sum_{u\in\Uc^{(b)}}\gamma_{u,b}\leq \Gamma_b\;\forall\;b\in\Bc_1';\\
 R_u\define R_{u,1}\theta_{u,1}+ R_{u,b}\gamma_{u,b}\in [R^{\rm min}_u, R^{\rm max}_u],\;\forall\;u\in\Uc^{(b)},b\in\Bc_1';
\end{aligned}}\end{equation}} \vspace{-1cm}}     \end{figure*}
which we assume to be feasible. 
 Note that without loss of generality we can assume that each pico TP is resource limited, i.e., $\sum_{u\in\Uc^{(b)}}\frac{R^{\rm max}_u}{R_{u,b}}>\Gamma_b$. This is because otherwise we can simply assign maximum possible resource from TP $b$ to each user in $\Uc^{(b)}$ and remove those from further consideration. We will use the term {\em slack} to denote the resource assigned to a user in excess of its minimum rate requirement. For convenience, for each user $k\in\Uc^{(b)},b\in\Bc_1'$
 we let $R_k=R_{k,1}\theta_{k,1}+ R_{k,b}\gamma_{k,b}$ and supress the dependence of $R_k$ on $\theta_{k,1},\gamma_{k,b}$.
To analyze (\ref{eq:Bigoriginalwsrp}) we offer the following result that can be derived by carefully manipulating the K.K.T. conditions.
\begin{propo}\label{propocond}
The following conditions must be satisfied by any optimal solution of (\ref{eq:Bigoriginalwsrp}):
\begin{itemize}
\item For any two distinct users associated with any TP, $k,j\in\Uc^{(b)},\;b\in\Bc_1'$, such that
 $\frac{R_{k,b}}{R_{k,1}}>\frac{R_{j,b}}{R_{j,1}}$, we must have
\begin{eqnarray}
\theta_{k,1}>0\Rightarrow \gamma_{j,b}=0
\end{eqnarray}
\item {\em Slack ordering:} For any two distinct users associated with any TP, $k,j\in\Uc^{(b)},\;b\in\Bc_1'$, such that
 $w_kR_{k,b}>w_jR_{j,b}$, we must have
\begin{eqnarray}
R_j>R_j^{\rm min}\;\&\;\gamma_{j,b}>0\Rightarrow R_k = R^{\rm max}_k.
\end{eqnarray}
Similarly, for any two distinct users, $k,j\in\Uc'$, such that
 $w_kR_{k,1}>w_jR_{j,1}$, we must have
\begin{eqnarray}
R_j>R_j^{\rm min}\;\&\;\theta_{j,1}>0\Rightarrow R_k = R^{\rm max}_k.
\end{eqnarray}
\item For any user $k\in\Uc^{(b)},\;b\in\Bc_1'$, such that $\gamma_{k,b}>0$, if there exists any other user $j\in\Uc'$ with
 $\theta_{j,1}>0$ and $R_j>R_j^{\rm min}$ , we must have
\begin{eqnarray}
\frac{R_{k,b}}{R_{k,1}}\geq \frac{\max_{j\in\Uc^{(b)}\setminus k :R_j<R_j^{\rm max}}\{w_jR_{j,b}\}}{\min_{j\in\Uc'\setminus k:\theta_{j,1}>0 \;\&\;R_j>R_j^{\rm min}}\{w_jR_{j,1}\}}.
\end{eqnarray}
Similarly, for any user $k\in\Uc^{(b)},\;b\in\Bc_1'$, such that $\theta_{k,1}>0$, if there exists any other user $j\in\Uc^{(b)}$ with
 $\gamma_{j,b}>0$ and $R_j>R_j^{\rm min}$ , we must have
\begin{eqnarray}
\frac{R_{k,b}}{R_{k,1}} \leq \frac{\min_{j\in\Uc^{(b)}\setminus k :\gamma_{j,b}>0\;\&\;R_j>R_j^{\rm min}}\{w_jR_{j,b}\}}{\max_{j\in\Uc'\setminus k: R_j<R_j^{\rm max}}\{w_jR_{j,1}\}}.
\end{eqnarray}
\end{itemize}
\end{propo}

Letting $\hat{O}(\Gamma,\{\Gamma_b\})$ denote the optimal value of (\ref{eq:Bigoriginalwsrp}) for the given budgets,
 we embark to solve   (\ref{eq:Bigoriginalwsrp}) and characterize $\hat{O}(\Gamma,\{\Gamma_b\})$. 
Towards that end, without loss of generality, in this section we assume a labeling of user indices such that for any two users $k,j\in\Uc^{(b)}$ for any $b\in\Bc_1'$, $k<j\Rightarrow \frac{R_{k,b}}{R_{k,1}}\geq \frac{R_{j,b}}{R_{j,1}}$. Further, without loss of generality, the user indices in  $\Uc^{(b)}$ for each $b\in\Bc_1'$, are assumed to be consecutive.
 Then, upon appyling primal decomposition on (\ref{eq:Bigoriginalwsrp}), we see  that if we fix the share of the Macro resource that can be used by each TP $b\in\Bc_1'$ as $Z_b$, where
 $\sum_{b\in\Bc_1'}Z_b=\Gamma$,  (\ref{eq:Bigoriginalwsrp}) decouples into $|\Bc_1'|$ sub-problems. In particular,
 the problem at hand for TP $b$ is given by (\ref{eq:Bigoriginalwsrpb}).
 \begin{figure*}{{\begin{equation}\boxed{\begin{aligned}\label{eq:Bigoriginalwsrpb}
   \max_{\gamma_{u,b},\theta_{u,1}\in[0,1]\;\forall\;u\in\Uc^{(b)}}\left\{\sum_{u\in\Uc^{(b)}}\left(w_u R_{u,1}\theta_{u,1}+ w_uR_{u,b}\gamma_{u,b}\right)\right\}\\
 {\rm s.t. }  \sum_{u\in\Uc^{(b)}}\theta_{u,1}\leq Z_b\;\&\; \sum_{u\in\Uc^{(b)}}\gamma_{u,b}\leq \Gamma_b;\\
 R_u \in [R^{\rm min}_u, R^{\rm max}_u],\;\forall\;u\in\Uc^{(b)};
\end{aligned}}\end{equation}} \vspace{-1cm} }    \end{figure*}
Let $\hat{O}(Z_b,\Gamma_b,b)$ denote the optimal solution of (\ref{eq:Bigoriginalwsrpb}) for each TP $b\in\Bc_1'$ so that the original problem in (\ref{eq:Bigoriginalwsrp}) can be expressed as
\begin{eqnarray}\label{eq:wsrpp}
\max_{\{Z_b\in\Reals_+\}:\sum_{b\in\Bc_1'}Z_b\leq \Gamma}\left\{  \sum_{b\in\Bc_1'}\hat{O}(Z_b,\Gamma_b,b)            \right\}
\end{eqnarray}
   A straightforward approach  to determine the optimal Macro resource share among the TPs is to optimize $\{Z_b\}$ using the generic subgradient method. However, we will show that exploiting the structure of the problem at hand leads directly to a very simple algorithm.
First, let us define the function, $\bar{h}: \Reals_+\times\Bc_1'\to\Reals_+$, such that  $\bar{h}(\Gamma_b,b)$  for any given TP $b$ and corresponding budget, $\Gamma_b$, yields the minimum Macro resource needed (in addition to the available budget $\Gamma_b$ for the pico TP $b$) accommodate the minimum rates of all users in $\Uc^{(b)}$.    In particular, we can invoke Proposition \ref{propocond} to explicitly detail $\bar{h}(\Gamma_b,b)$ after   recalling the labelling we have adopted, as
\begin{eqnarray}\label{eq:tildeG}
\nonumber \bar{h}(\Gamma_b,b)=  \left\{\begin{array}{lc}
\frac{R^{\rm min}_{\tilde{k}+1}-\Xi_bR_{\tilde{k}+1,b}}{R_{\tilde{k}+1,1}} +    \sum_{j\in\Uc^{(b)}:j> \tilde{k}+1}\frac{R_{j}^{\rm min}}{R_{j,1}},   &  \tilde{k}+1\in\Uc^{(b)} \\
0,   & {\rm Else} \end{array} \right.\\
 \tilde{k}=\max\left\{k: k\in\Uc^{(b)} \;\&\; \sum_{j\in\Uc^{(b)}:j\leq k}\frac{R_{j}^{\rm min}}{R_{j,b}} \leq\Gamma_b\right\},\;
\Xi_b = \Gamma_b-\sum_{j\in\Uc^{(b)}:j\leq \tilde{k}}\frac{R_{j}^{\rm min}}{R_{j,b}}
\end{eqnarray}
In (\ref{eq:tildeG}) we use the convention that $\tilde{k}+1$ reurns the user with the lowest index in $\Uc^{(b)}$ whenever
 $\tilde{k}$ is null on account of $\sum_{j\in\Uc^{(b)}:j\leq k}\frac{R_{j}^{\rm min}}{R_{j,b}} >\Gamma_b$ for all  $k\in\Uc^{(b)}$.
In a similar manner we define $\hul:\Reals_+\times\Bc_1'\to\Reals_+$, such that $\hul(Z_b,b)$  for any given TP $b$ and a given Macro  budget, $Z_b$, yields the additional minimum resource needed by TP $b$ to accommodate the minimum rates of all users in $\Uc^{(b)}$. Again invoking Proposition \ref{propocond} we can explicitly detail $\hul(Z_b,b)$, as
\begin{eqnarray}
\nonumber \hul(Z_b,b) =  \left\{\begin{array}{lc}
\frac{R^{\rm min}_{\breve{k}-1}-\tilde{Z}_bR_{\breve{k}-1,1}}{R_{\breve{k}-1,b}} +    \sum_{j\in\Uc^{(b)}:j< \breve{k}-1}\frac{R_{j}^{\rm min}}{R_{j,b}},   &  \breve{k}-1\in\Uc^{(b)} \\
0,   & {\rm Else} \end{array} \right.\\
 \breve{k}=\min\left\{k: k\in\Uc^{(b)} \;\&\; \sum_{j\in\Uc^{(b)}:j\geq k}\frac{R_{j}^{\rm min}}{R_{j,1}} \leq Z_b\right\},\;
\tilde{Z}_b = Z_b-\sum_{j\in\Uc^{(b)}:j\geq \breve{k}}\frac{R_{j}^{\rm min}}{R_{j,1}}
\end{eqnarray}
For any given $\Gamma_b$, we let $S(Z_b,\Gamma_b,b),\;b\in\Bc_1',Z_b\geq \bar{h}(\Gamma_b,b)$ denote the slope of the  function $\hat{O}(Z_b,\Gamma_b,b)$ at $Z_b$. In particular, 
$S(Z_b,\Gamma_b,b) = \lim_{\delta\to 0_+}\frac{\hat{O}(Z_b+\delta,\Gamma_b,b)- \hat{O}(Z_b,\Gamma_b,b)}{\delta}$. 
Henceforth, without loss of generality, we assume $\hul(1,b)\leq 1\;\&\;\bar{h}(1,b)\leq 1,\;\forall\;b\in\Bc_1'$.
\begin{propo}\label{prop:charsub}
For any fixed $\Gamma_b\geq \hul(1,b)$, $\hat{O}(Z_b,\Gamma_b,b)$ is continuous, non-decreasing, piecewise linear and concave in $Z_b\in[\bar{h}(\Gamma_b,b),1]$.
 For any fixed  $Z_b\geq \bar{h}(1,b)$, $\hat{O}(Z_b,\Gamma_b,b)$ is continuous, non-decreasing, piecewise linear and concave in $\Gamma_b\in[\hul(Z_b,b),1]$.
\end{propo}
 \proof We only prove the first claim since proof for the second one follows along similar lines. The continuity and non-decreasing properties are straightforward to verify.
It can be shown that the conditions stated in Proposition \ref{propocond}  provide necessary and sufficient conditions  to determine an  optimal set of allocation fractions for the problem in (\ref{eq:Bigoriginalwsrpb}).
To verify the other two properties, we start at  $Z_b=\bar{h}(\Gamma_b,b)$. Then, if $\bar{h}(\Gamma_b,b)=0$, the slack at the pico TP   $b$, $\Gamma_b -  \sum_{j\in\Uc^{(b)}}\frac{R_{j}^{\rm min}}{R_{j,b}}$ must be distributed among users in the decreasing order  $\{w_kR_{k,b}\}_{k\in\Uc^{(b)}}$ subject to their respective maximum rate limits (cf. slack ordering in Proposition \ref{propocond}). On the other hand,
 when  $\bar{h}(\Gamma_b,b)>0$ there is no slack at the pico TP for this  $Z_b$.
 The next key observation we use is the one in Proposition \ref{propocond} pertaining to the order in which macro resources are assigned to users in $\Uc^{(b)}$. Following our labelling, we see that when $Z_b=\bar{h}(\Gamma_b,b)>0$ either user $\tilde{k}+1$ (when $ \Xi_b>0$) or user $\tilde{k}$ (when $ \Xi_b=0$) is the  user with the largest index in $\Uc^{(b)}$ to be assigned a positive resource by TP $b$. Let user $k'$ be this user so that users $k\in\Uc^{(b)}:k<k'$ are assigned resource only by TP $b$  
 at this $Z_b=\bar{h}(\Gamma_b,b)$. The slope of $\hat{O}(Z_b,\Gamma_b,b)$ can be determined as
 $S(\bar{h}(\Gamma_b,b),\Gamma_b,b)=\max\left\{\max\left\{\frac{w_kR_{k,b}R_{k',1}}{R_{k',b}}: k\in\Uc^{(b)}\;\&\;k< k'\right\}, \max\{ w_kR_{k,1}: k\in\Uc^{(b)}\;\&\;k\geq k'\}  \right\}$. Then, as $Z_b$ is increased to $Z_b+\delta$, for any arbitrarily small $\delta>0$, the slack is put  to the user yielding  the  slope $S(\bar{h}(\Gamma_b,b),\Gamma_b,b)$ (i.e., offering the maximum bang-per-buck). If such a user is some $\hat{k} \geq k'$ then the additional available Macro resource, $\delta$, is directly assigned as slack to it. Otherwise, the additional available Macro resource is first assigned to user $k'$,
 which frees up resource $\frac{\delta R_{k',1}}{R_{k',b}}$ at the pico TP $b$ (while maintaining the minimum rate of user $k'$). This freed up pico resource is assigned as slack to the user $\hat{k}$ yielding  the slope $S(\bar{h}(\Gamma_b,b),\Gamma_b,b)$.
  As $\delta$ is increased the slack is continuously assigned to the user $\hat{k}$ till the slope changes and that user is not the one yielding the maximum bang-per-buck. This happens either if user $\hat{k}$ attains its maximum rate upon which it is removed from the  candidate list of users which can be assigned slack, or if user $k'$ is no longer assigned any resource  by pico TP $b$. In the former case,  the slope changes to $S(Z_b,\Gamma_b,b)=\max\left\{\max\left\{\frac{w_kR_{k,b}R_{k',1}}{R_{k',b}}: k\in\Uc^{(b)}\setminus\hat{k}\;\&\;k< k'\right\}, \max\{ w_kR_{k,1}: k\in\Uc^{(b)}\setminus\hat{k}\;\&\;k\geq k'\}  \right\}$, whereas in the latter case
 the slope changes to $S(Z_b,\Gamma_b,b)=\max\left\{\max\left\{\frac{w_kR_{k,b}R_{k'-1,1}}{R_{k'-1,b}}: k\in\Uc^{(b)}\;\&\;k< k'-1\right\}, \max\{ w_kR_{k,1}: k\in\Uc^{(b)}\;\&\;k\geq k'-1\}  \right\}$.  Thenceforth additional Macro resources are assigned as slack to the user yielding the new slope and the process continues.  Note that at every change the slope decreases because either users are removed from candidate list or the gain term multiplying the weight of each user served exclusively by the pico reduces.  This demonstrates the piecewise linearity and concavity.
 The same arguments can be applied when  $\bar{h}(\Gamma_b,b)=0$. In particular, we begin at  $Z_b=\bar{h}(\Gamma_b,b)=0$ after  determining user $k'$   that has the highest index among those that have been assigned a positive resource by the pico TP and after removing users that have achieved their maximum rates from the candidate pool. The subsequent process proceeds as before and we can deduce the piecewise linearity and concavity.
\endproof
\begin{corollary}\label{coro:charsub}
For any fixed $\Gamma_b\geq \hul(1,b)$, $\hat{O}(Z_b,\Gamma_b,b)$ can be computed as
\begin{eqnarray}\label{eq:corosub}
\hat{O}(Z_b,\Gamma_b,b) = \sum_{k\in\Uc^{(b)}}w_kR_k^{\rm min}  + H(\Gamma_b,b)+  \int_{\bar{h}(\Gamma_b,b)}^{Z_b} S(z_b,\Gamma_b,b)dz_b,\;\forall\; Z_b\in[\bar{h}(\Gamma_b,b),1],
\end{eqnarray}
where $H(\Gamma_b,b)=0$ whenever $\bar{h}(\Gamma_b,b)>0$ and when $\bar{h}(\Gamma_b,b)=0$, it yields the  weighted sum rate obtained by distributing the excess pico resource as slack among users in $\Uc^{(b)}$
 in the    decreasing order  $\{w_kR_{k,b}\}$ subject to their respective maximum rate limits.
\end{corollary}


We now propose Algorithm \ref{algo:wsrallocopt} to determine   Macro allocations that optimize (\ref{eq:wsrpp}) (and so  (\ref{eq:Bigoriginalwsrp})). 
\begin{table}
\caption{{\bf Algorithm I: WSR optimal allocation fractions}}\label{algo:wsrallocopt}
\begin{algorithmic}[1]
\STATE Initialize with  $\Bc_1', \Uc^{(b)},\;b\in\Bc_1'$. Set  $Z_b=\bar{h}(\Gamma_b,b),\;\forall\;b\in\Bc_1'$ and $C=\Gamma-\sum_{b\in\Bc_1'}Z_b$.
\STATE For each pico TP $b\in\Bc_1'$, if $Z_b=0$ then distribute the slack at that pico,  $\Gamma_b -  \sum_{j\in\Uc^{(b)}}\frac{R_{j}^{\rm min}}{R_{j,b}}$,    among users in $\Uc^{(b)}$. 
\STATE \textbf{Repeat}
\STATE Determine $\hat{b} = \arg\max_{b\in\Bc_1'}\{S( Z_b,\Gamma_b,b)\} $
\STATE Determine $\hat{\Delta} =  \sup \{\Delta\in\Reals_+: S( Z_{\hat{b}}+\Delta,\Gamma_{\hat{b}},\hat{b}) = S( Z_{\hat{b}},\Gamma_{\hat{b}},\hat{b}) \} $.
 \STATE  Increment $Z_{\hat{b}}= Z_{\hat{b}} + \min\{C,\hat{\Delta}\}$ and update $C = \max\{0,C - \hat{\Delta}\}$.
\STATE \textbf{Until} $C=0$.
\STATE Output $\{Z_b\},\forall\;b\in\Bc_1'$, the corresponding  allocation fractions $\{\theta_{u,1},\gamma_{u,b}\},\;\forall\;u\in\Uc^{(b)},b\in\Bc_1'$.
\end{algorithmic}  
\vspace{-1cm}\end{table}


\begin{propo}\label{prop:optchar}
The optimal solution to (\ref{eq:Bigoriginalwsrp})   can be determined  using Algorithm \ref{algo:wsrallocopt} whenever the necessary and sufficient condition for feasibility, $\sum_{b\in\Bc'}\bar{h}(\Gamma_b,b)\leq 1$, holds.
For any fixed budgets $\{\Gamma_b\}\;\forall\;b\in\Bc_1'$ satisfying the feasibility condition, $\hat{O}(\Gamma,\{\Gamma_b\})$ is continuous, non-decreasing, piecewise linear and concave in $\Gamma\in[\sum_{b\in\Bc_1'}\bar{h}(\Gamma_b,b),1]$.
\end{propo}
\proof First note that using Proposition \ref{prop:charsub} with (\ref{eq:wsrpp}), we see that we are maximizing $|\Bc_1'|$ piecewise linear concave functions subject to a linear budget constraint. Notice that in Algorithm \ref{algo:wsrallocopt} we always choose the pico TP yielding the highest slope and assign it as much resource as possible till the point that maximal slope changes. This greedy strategy is optimal for the problem at hand because: (i) each slope curve  $S(Z_b,\Gamma_b,b), Z_b\geq \bar{h}(\Gamma_b,b)$ is  a piecewise constant function in $Z_b$ and (ii) any Macro resource assigned to any TP $b'$ has no influence on the slope curve of any other TP $b\neq b',b\in\Bc_1'$. More formally, (\ref{eq:wsrpp}) can be shown to equivalent to the maximization of a modular function subject to  a cardinality constraint for which the greedy strategy is optimal.  Next, the claimed properties  of $\hat{O}(\Gamma,\{\Gamma_b\})$  directly follow from the facts in Proposition \ref{prop:charsub}   that  in  (\ref{eq:wsrpp}) each $\hat{O}(Z_b,\Gamma_b,b)$ is continuous, non-decreasing, piecewise linear and concave in $Z_b$.
\endproof
Let $\Gammab=[\Gamma_b]_{b\in\Bc_1'}$ denote any  vector of all pico budgets and   let
 $S(Z,\Gammab)$ denote the slope curve of $\hat{O}(Z,\Gammab)=\hat{O}(Z,\{\Gamma_b\})$  for  $Z\geq\sum_{b\in\Bc_1'}\bar{h}(\Gamma_b,b)$, which from Proposition \ref{prop:optchar}  we know to be piecewise constant and non-increasing in $Z$. We then have the following corollary.
\begin{corollary}
\begin{eqnarray}\label{eq:corosubmain}
\hat{O}(\Gamma,\Gammab) = \sum_{b\in\Bc_1'}\sum_{k\in\Uc^{(b)}}w_kR_k^{\rm min}  + \sum_{b\in\Bc_1'}H(\Gamma_b,b)+  \int_{\sum_{b\in\Bc_1'}\bar{h}(\Gamma_b,b)}^{\Gamma} S(z,\Gammab)dz,\;\forall\; \Gamma \in\left[\sum_{b\in\Bc_1'}\bar{h}(\Gamma_b,b),1\right].
\end{eqnarray}
\end{corollary}
To illustrate our results, in Fig. \ref{fig:DCwsr} we consider a macro TP serving $|\Uc'|=30$ users, with $|\Bc_1|=10$ pico nodes assigned to it and where each user is associated with the pico in $\Bc_1$ from which it sees the strongest signal strength. We obtained the user peak rates by emulating a realistic deployment (the details are defered to the simulation results section) and imposed no maximum rate limits (i.e., each $R^{\rm max}_u=\infty$). A unit budget at each pico was assumed and  the minimum rate of each user was chosen to be a scalar times its peak rate from the macro TP (with the scalar being identical for each user).  We considered several values for this scalar and in each case  plot $\hat{O}(\Gamma,\{1\}),\Gamma \in [\sum_{b\in\Bc'_1}\bar{h}(1,b),1]$ (computed using Algorithm \ref{algo:wsrallocopt}).  As predicted by Proposition \ref{prop:optchar}, each curve is non-decreasing, piece-wise linear and concave and we verified that the obtained value matches the one obtained by solving   (\ref{eq:Bigoriginalwsrp}) via a generic LP solver. Notice  as the minimum rate requirements become more stringent, more macro resource is needed to satisfy them and the optimized utility value decreases.
Before offering our next key result, which is proved in the appendix, we introduce some notation. For any two pico TPs $b_1,b_2\in\Bc_1'$, we let
 $\eb_{b_1},\eb_{b_2}$ define $|\Bc_1'|\times 1$ unit vectors that have a zero on all their entries except those corresponding to $b_1,b_2$, respectively, which  are both one.
\begin{propo}\label{prop:wsrSs}
For any non-negative scalars $\delta,\tilde{\delta}, \delta_{b_1},\tilde{\delta}_{b_2}$ and budgets $\Gamma,\Gammab$ such that
 $\Gamma\geq \sum_{b\in\Bc_1'}\bar{h}(\Gamma_b,b)$,  we have that
 \begin{eqnarray}\label{eq:subnomax}
 \hat{O}(\Gamma,\Gammab) - \hat{O}(\Gamma+\delta,\Gammab+\delta_{b_1}\eb_{b_1})\leq  \hat{O}(\Gamma+\tilde{\delta}, \Gammab +    \tilde{\delta}_{b_2}\eb_{b_2}) - \hat{O}(\Gamma+\tilde{\delta}+\delta,  \Gammab + \delta_{b_1}\eb_{b_1} +  \tilde{\delta}_{b_2}\eb_{b_2}  ).\;
\end{eqnarray}
 \end{propo}


 \subsection{Optimal allocation fractions of (\ref{eq:Bigoriginal})}\label{sec:charpf}
As before we obtain the sets $\Uc',\Bc_1'$ from the given assoication and  let $N_b=|\Uc^{(b)}|$ denote the cardinality or the number of users associated with TP $b\in\Bc_1'$.  
Consider the PF system utility optimization problem (restricted to the user pool $\Uc'$) for the the given user association, given in
 (\ref{eq:Bigoriginalp2}).
\begin{figure*}{{\begin{equation}\boxed{\begin{aligned}\label{eq:Bigoriginalp2}
   \max_{\gamma_{u,b},\theta_{u,1}\in[0,1]\;\forall\;u\in\Uc',b\in\Bc_1'}\left\{\sum_{u\in\Uc'}\sum_{b\in\Bc_1'}x_{u,b}\ln\left( R_{u,1}\theta_{u,1}+ R_{u,b}\gamma_{u,b}\right)\right\}\\
 {\rm s.t. }
 \sum_{u\in\Uc'}\theta_{u,1}\leq 1\;\&\; \sum_{u\in\Uc'}\gamma_{u,b}\leq 1\;\forall\;b\in\Bc_1'.
\end{aligned}}\end{equation}}\vspace{-1cm}}      \end{figure*}
Note that (\ref{eq:Bigoriginalp2}) is a purely continuous optimization problem. 
Next, for each $b\in\Bc_1'$ define
 $\mu_{1,b}=\min_{u\in\Uc^{(b)}} \{ \frac{R_{u,1}}{R_{u,b}}\}$. Similary,
let $ \mu_{k,b}, k\in\{2,\cdots,N_b\}$ denote that $k^{th}$ smallest  ratio in the set $\{ \frac{R_{u,1}}{R_{u,b}}\}_{u\in\Uc^{(b)}}$ and recall that these ratios are all strictly positive and distinct. Then, defining $\mu_{N_b+1,b}=\infty$, we have $0 < \mu_{1,b} < \mu_{2,b} < \cdots <\mu_{N_b,b}<\mu_{N_b+1,b}=\infty$.
Next, we define a function $h:\Reals_{++}\times \Bc_1'\to\Reals_+$ as
\begin{eqnarray}
h(\lambda,b) =  \left\{\begin{array}{lc}
\frac{1}{\mu_{m,b}},   & m=1,\cdots,N_b:\lambda\in((m-1)\mu_{m,b},m\mu_{m,b}) \\
\frac{m-1}{\lambda},   &  m=2,\cdots,N_b+1:\lambda\in[(m-1)\mu_{m-1,b},(m-1)\mu_{m,b}] \end{array} \right.
\end{eqnarray}
Similarly, we define  function $g:\Reals_{++}\times \Bc_1'\to\Reals_+$ such that, $g(\lambda,b),\forall\;\lambda>0,b\in\Bc'_1$ equals
\begin{eqnarray}
  \left\{\begin{array}{lc}
\sum_{j=m}^{N_b}\ln(\mu_{j,b}/\lambda) + (m-1)\ln(\mu_{m,b}/\lambda),   &  m=1,\cdots,N_b:\lambda\in((m-1)\mu_{m,b},m\mu_{m,b})\\
-(m-1)\ln(m-1) + \sum_{q=m}^{N_b}\ln(\mu_{q,b}/\lambda),   & m=2,\cdots,N_b+1:\lambda\in[(m-1)\mu_{m-1,b},(m-1)\mu_{m,b}] \end{array} \right.
\end{eqnarray}
The following result is proved in the appendix. 
\begin{theorem}\label{thm:pfalloc}
The optimal objective value of (\ref{eq:Bigoriginalp2}) is given by
\begin{eqnarray}
\sum_{b\in\Bc_1'}\left(g(\hat{\lambda},b)+\sum_{k\in\Uc^{(b)}}\ln(R_{k,b})\right),
\end{eqnarray}
 where $\hat{\lambda}\in (0,\infty)$ is the unique solution to the relation
\begin{eqnarray}\label{eq:lrel}
 1 + \sum_{b\in\Bc_1'}h(\lambda,b) = \sum_{b\in\Bc_1'}N_b/\lambda,
\end{eqnarray}
which can be determined via bisection search.
\end{theorem}
\begin{corollary}\label{coro:pf}
Suppose that the optimal $\hat{\lambda}$ satisfying (\ref{eq:lrel}) is given. Then, if
$ \hat{\lambda}\in((m-1)\mu_{m,b},m\mu_{m,b})$ for some $m=1,\cdots,N_b$, the
 optimal solution comprises of
 assigning an identical resource share $\gamma_{k,b}=\mu_{m,b}/\hat{\lambda}$ with $\theta_{k,1}=0$ for all users $k\in\Uc^{(b)}:\frac{R_{k,1}}{R_{k,b}}<\mu_{m,b}$, whereas all users $k\in\Uc^{(b)}:\frac{R_{k,1}}{R_{k,b}}>\mu_{m,b}$ are assigned $\theta_{k,1}=1/\hat{\lambda}$ with $\gamma_{k,b}=0$. The user $k\in\Uc^{(b)}:\frac{R_{k,1}}{R_{k,b}}=\mu_{m,b}$ is assigned $\theta_{k,1}=m/\hat{\lambda}-1/\mu_{m,b}$ with $\gamma_{k,b}=1 - (m-1)\mu_{m,b}/\hat{\lambda}$.
 On the other hand, if
$ \hat{\lambda}\in[(m-1)\mu_{m-1,b},(m-1)\mu_{m,b}]$ for some $m=2,\cdots,N_b+1$, the
 optimal solution comprises of
 assigning an identical resource share $\gamma_{k,b}=1/(m-1)$ with $\theta_{k,1}=0$ for all users $k\in\Uc^{(b)}:\frac{R_{k,1}}{R_{k,b}}<\mu_{m,b}$, whereas all users $k\in\Uc^{(b)}:\frac{R_{k,1}}{R_{k,b}}\geq \mu_{m,b}$ are assigned $\theta_{k,1}=1/\hat{\lambda}$ with $\gamma_{k,b}=0$.
\end{corollary}

We next introduce another useful result that will be invoked  to establish the performance guarantee of an algorithm proposed later for (\ref{eq:Bigoriginal}) in the sequel. Towards that end,
 we introduce the problem in (\ref{eq:Optorth}) where we recall that $\{x_{u,b}\}_{u\in\Uc',b\in\Bc'_1}$ are given.
 \begin{figure*}{{\begin{equation}\boxed{\begin{aligned}\label{eq:Optorth}
   \max_{\tilde{z}_{u,1}\in\{0,1\}\;\forall\;u\in\Uc'}\left\{\sum_{u\in\Uc'}\left(\tilde{z}_{u,1}\ln\left( R_{u,1}\right) +
   \sum_{b\in\Bc'_1}(1-\tilde{z}_{u,1})x_{u,b}\ln\left( R_{u,b}\right) \right) - \;\;\;\;\;\;\;\;\;\;\;\;\;\;\;\;\;\right.\\\left. \left(\left(\sum_{k\in\Uc'}\tilde{z}_{k,1}\right)\ln\left(\sum_{k\in\Uc'}\tilde{z}_{k,1}\right) + \sum_{b\in\Bc'_1}\left(\sum_{k\in\Uc'}(1-\tilde{z}_{k,1})x_{k,b}\right)\ln\left(\sum_{k\in\Uc'}(1-\tilde{z}_{k,1})x_{k,b}\right)\right)    \right\}
\end{aligned}}\end{equation}}\vspace{-1cm}}      \end{figure*}%
 \begin{propo}\label{prop:solnsimpO}
 The optimal solution determined from (\ref{eq:Optorth}) yields  an objective value for (\ref{eq:Bigoriginalp2}) that is no less than the optimal objective value of (\ref{eq:Bigoriginalp2}) minus $\min\{|\Bc'_1|,|\Uc'|\}\ln(2)$.
 \end{propo}
 \proof  We first note that the problem in (\ref{eq:Optorth}) is indeed equivalent to (\ref{eq:Bigoriginalp2}) with the additional restriction that $\theta_{u,1}\gamma_{u,b}=0,\;\forall\;u\in\Uc^{(b)},b\in\Bc'_1$. Further invoking the result in Corollary \ref{coro:pf} that in an optimal solution of (\ref{eq:Bigoriginalp2}), for each TP $b\in\Bc'_1$ at-most one user in $\Uc^{(b)}$ is assigned resources by both the macro TP and pico TP $b$. The remaining users in $\Uc^{(b)}$ are all assigned resource by either the macro TP or by pico TP $b$. Suppose that user is $\hat{u}_b$ which is assigned resource $\hat{\theta}_{\hat{u}_b,1}$ by the macro and resource
$ \hat{\gamma}_{\hat{u}_b,b}$ by pico  TP $b$.
  Then, for that user we can bound the rate as
 {{{ \begin{eqnarray*}
 \ln\left( R_{\hat{u}_b,1}\hat{\theta}_{\hat{u}_b,1}   + R_{\hat{u}_b,b}\hat{\gamma}_{\hat{u}_b,b} \right)
 \leq \ln(2)+\max\left\{ \ln\left( R_{\hat{u}_b,1}\hat{\theta}_{\hat{u}_b,1}\right), \ln\left(R_{\hat{u}_b,b}\hat{\gamma}_{\hat{u}_b,b}\right)    \right\},
    \end{eqnarray*}}}
Considering all TPs in $\Bc_1'$ we get the desired result. \endproof
\section{Approximation Algorithms}
We are now ready to propose approximation algorithms for the problems in (\ref{eq:Bigoriginalwsr2}) and (\ref{eq:Bigoriginal}).
We begin with the WSR maximization problem in (\ref{eq:Bigoriginalwsr2}).
Let us define a ground set $\Omegaul=\{(u,b),u\in\Uc,b\in\Pc\}$ where $(u,b)$ conveys the association of user $u$ with pico TP $b$. The tuple also implicitly indicates the association of $u$ to the Macro TP $m$ where $b\in\Bc_m$.  Without loss of generality we suppose that only a tuple $(u,b)$ for any $u\in\Uc\;\&\;b\in\Bc_m,m\in\Mc$ for which
 $R_{u,m}+R_{u,b}\geq R^{\rm min}_u$ is included in $\Omegaul$. This is because any tuple not satisfying this assumtion will never be selected as its minimum rate cannot be met even when the assigned macro and the  pico TPs fully allocate their resources to that user.
 Let $\Omegaul^{(m)}=\{(u,b)\in\Omegaul: b\in\Bc_m\}$ denote all possible  associations to any pico TP in $\Bc_m$, the set of pico TPs assigned  to macro TP $m$, and let   $\Omegaul_{(u')}=\{(u,b)\in\Omegaul: u=u'\}$ denote all possible associations of a user $u'\in\Uc$. Define a family of sets
$\Iulk$ as
 as the one which includes each subset of $\Omegaul$ such that the tuples in that subset have mutually distinct users. Formally,
 $\Aulc\subseteq\Omegaul: |\Aulc\cap\Omegaul_{(u)}|\leq 1\;\forall\; u\in\Uc \Leftrightarrow \Aulc\in\Iulk.$ 
Further, define a family, $\Julk$, contained in $\Iulk$ that comprises of each member of $\Iulk$ for which  (\ref{eq:Bigoriginalwsr2}) is feasible. Using the definitions given in the appendix, we see that while $\Iulk$ defines a matroid, $\Julk$ is   a downward closed family but  need not satisfy the exchange property and hence need not define a matroid.
 Next, we define a non-negative set function on $\Julk$, $f^{\rm wsr}:\Julk\to \Reals_+$ such that it is normalized, i.e.,
 $f(\phi)=0$,  and for any non-empty set $\Gulc\in\Julk$, we have
\begin{eqnarray}\label{eq:funcwsr}
f^{\rm wsr}(\Gulc) = \sum_{m\in\Mc} f^{\rm wsr}_m(\Gulc\cap\Omegaul^{(m)}).
\end{eqnarray}
Each $f^{\rm wsr}_m:\Julk^{(m)}\to \Reals_+$ in (\ref{eq:funcwsr}) is a normalized non-negative set function that is defined on the family
 $\Julk^{(m)}$ which comprises of each member of $\Julk$ that is contained in $\Omegaul^{(m)}$,
as follows. For any set $\Aulc\in\Julk^{(m)}$, we define
 $f^{\rm wsr}_m(\Aulc) = \hat{O}(1,{\bf 1})$, where $\hat{O}(1,{\bf 1})$ is computed as described in Algorithm \ref{algo:wsrallocopt} in Section \ref{sec:charwsr} for the macro TP $m$ and the set of pico TPs $\Bc_m$ assigned to it, using unit budgets and the given association in $\Aulc$.  We recall that a simple necessary and sufficient condition to determine feasibility of the minimum rates for the given association and budgets is provided in Proposition
  \ref{prop:optchar}.
With these definitions in hand,
can re-formulate the problem in (\ref{eq:Bigoriginalwsr2}) as the following constrained set function maximization problem.
 \begin{eqnarray}
 \max_{  \Gulc\in\Julk}\{ f^{\rm wsr}(\Gulc)\}
 \end{eqnarray}
We offer our first  main result that characterizes $f^{\rm wsr}(.)$.
\begin{theorem}\label{thm:wsrsubmod}
The set function $f^{\rm wsr}(.)$ is a normalized non-negative submodular set function and can be non-monotone.
\end{theorem}
\proof
The set function $f^{\rm wsr}(.)$  in (\ref{eq:funcwsr}) defined on the family $\Julk$ is normalized and non-negative by construction.  Due to the presence of minimum rate limits in (\ref{eq:Bigoriginalwsr2}) this function need not be monotone, i.e., there can exist members $\Ac,\Bc\in\Julk:\Ac\subseteq\Bc$ for which $  f^{\rm wsr}(\Ac)>f^{\rm wsr}(\Bc)$. Simultaneously, there can exist members $\Ac',\Bc'\in\Julk:\Ac'\subseteq\Bc'$ for which $  f^{\rm wsr}(\Ac')\leq f^{\rm wsr}(\Bc')$.
Then, to establish submodularity of $f^{\rm wsr}(.)$ on the family $\Julk$, it suffices  to show that each $f^{\rm wsr}_m(.)$ is submodular
 on the family $\Julk^{(m)}$.
 Without loss of generality, we consider macro TP $1$ and will prove    that
  forall $\Eulc\subseteq\Fulc\in\Julk^{(1)},\;(u_1,b_1)\in\Omegaul\setminus\Fulc:\Fulc\cup (u_1,b_1)\in \Julk^{(1)}$,
\begin{eqnarray}\label{eq:submod1}
f^{\rm wsr}_1(\Eulc\cup (u_1,b_1)) -  f^{\rm wsr}_1(\Eulc)
 \geq f^{\rm wsr}_1(\Fulc\cup (u_1,b_1)) -  f^{\rm wsr}_1(\Fulc).
\end{eqnarray}
Further, it suffices to prove (\ref{eq:submod1}) for $\Fulc=\Eulc\cup (u_2,b_2)$ so that $|\Fc|=|\Ec|+1$ and $(u_2,b_2)\in\Julk^{(1)}$.
Then, we evaluate $f^{\rm wsr}_1(\Fulc\cup (u_1,b_1))$ as described in Section \ref{sec:charwsr} and in the obtained optimal allocation fractions let the share of pico TP $b_1$ resource assigned to   user $u_1$ in tuple $(u_1,b_1)$ be $\delta_{b_1}$ and the share of macro TP resource assigned to that   user   be $\delta$.
 Similarly, let the share of pico TP $b_2$ resource assigned to   user $u_2$ in tuple $(u_2,b_2)$ be $\tilde{\delta}_{b_2}$ and the share of macro TP resource assigned to that  user  be $\tilde{\delta}$. Define $\Gamma=1-\delta-\tilde{\delta}$ and $\Gammab ={\bf 1}-\delta_{b_1}\eb_{b_1} - \tilde{\delta}_{b_2}\eb_{b_2}$.
Thus, we have that
\begin{eqnarray}\label{eq:subeq1}
  f^{\rm wsr}_1(\Fulc\cup (u_1,b_1)) =  \hat{O}(\Gamma, \Gammab) + w_{u_1}R_{u_1,b_1}\delta_{b_1} + 
w_{u_1}R_{u_1,1}\delta +
w_{u_2}R_{u_2,b_2}\tilde{\delta}_{b_2} + w_{u_2}R_{u_2,1}\tilde{\delta},
\end{eqnarray}
where  $\hat{O}(\Gamma, \Gammab)$ is evaluated for the tuples in $\Eulc$ under the budgets $\Gamma$ and $\Gammab$.
Further, we can  readily verify the relations in (\ref{eq:subineqs}).
\begin{figure*}\begin{eqnarray}\label{eq:subineqs}
\nonumber  f^{\rm wsr}_1(\Fulc) &\geq&   \hat{O}(\Gamma+\delta, \Gammab+\delta_{b_1}\eb_{b_1}) +
w_{u_2}R_{u_2,b_2}\tilde{\delta}_{b_2} + w_{u_2}R_{u_2,1}\tilde{\delta};\\
\nonumber f^{\rm wsr}_1(\Eulc) &=&   \hat{O}(\Gamma+\delta+\tilde{\delta}, \Gammab+\delta_{b_1}\eb_{b_1}+\tilde{\delta}_{b_2}\eb_{b_2}) \\
 f^{\rm wsr}_1(\Eulc\cup (u_1,b_1)) &\geq&  \hat{O}(\Gamma +\tilde{\delta}, \Gammab+\tilde{\delta}_{b_2}\eb_{b_2}) + w_{u_1}R_{u_1,b_1}\delta_{b_1} + w_{u_1}R_{u_1,1}\delta.\;\;\;\;\;
\end{eqnarray}\vspace{-1cm}\end{figure*}
Using (\ref{eq:subeq1}) and  (\ref{eq:subineqs}) in (\ref{eq:submod1}), it is now seen that a sufficient condition for  (\ref{eq:submod1}) to hold is for (\ref{eq:subnomax}) to be true. The latter is assured by Proposition \ref{prop:wsrSs}. \endproof 

We propose Algorithm \ref{algo:gels} referred to as the Greedy plus Enhanced Local Search (GELS) algorithm to optimize (\ref{eq:Bigoriginalwsr2}).  This algorithm is an adaptation (with a slight variation) of the one proposed in \cite{Lee:submod}    for non-monotone submodular set function maximization under a matroid constraint.
We note that the Algorithm \ref{algo:gels} can build a set using an optional  greedy stage. This set is further refined in the enhanced local search stage comprising of addition, deletion and swap operations. At the termination  of the LS stage we obtain the primary choice $\breve{\Gulc}$. Then, both   stages are repeated  over the complement set, $\Omegaul\setminus\breve{\Gulc}$ to generate an alternate choice, $\tilde{\Gulc}$. Finally, the  choice yielding the larger weighted sum rate utility among the primary and alternate choices is chosen. Regarding the slight variation alluded to above, we note that the direct adaptation would have initialized the Enhanced LS stage with the empty set or the singleton set yielding the highest weighted rate, since the LS stage also includes adding (or insertion) of elements.
However,  in our numerical simulations we saw that initializing the LS stage using the output of the greedy stage helps in reducing the run-time without performance degradation.
We proceed to derive performance guarantee for Algorithm \ref{algo:gels}.
Towards that end, we introduce an assumption pertaining to the feasibility of the minimum rates. We emphasize that this assumption is only for deriving a performance guarantee and is not needed for implementing the algorithm.\\
 $\bullet$ {\em Admission control assumption:  Each macro TP $m\in\Mc$ can itself simultaneously meet twice the minimum rates of all users $u$ that are present in at-least one tuple  $(u,b)\in\Omegaul$ for any $b\in\Bc_m$.}\\
We now offer the following result which holds even when the greedy stage is skipped and which assumes that  Algorithm \ref{algo:gels} is initialized with
 ${\rm MaxIter}=\infty$ and $\Delta=\frac{\epsilon}{C}$, where $\epsilon>0$ and  $C$ is a large enough constant that depends (polynomially) on the size
 of $\Omegaul$.
\begin{theorem}
 Algorithm \ref{algo:gels} yields a constant factor ($\frac{1}{4+\epsilon}$) approximation to (\ref{eq:Bigoriginalwsr2}) over all input instances for which the  admission control assumption holds.
\end{theorem}
\proof Let $\tilde{\Iulk}$ denote the family of sets obtained by taking the pairwise union of members of $\Iulk$. 
We define an extended set function as $\tilde{f}^{\rm wsr}(\Gulc)=\sum_{m\in\Mc}\tilde{ f}^{\rm wsr}_m(\Gulc\cap\Omegaul^{(m)}),\;\forall\;\Gulc\in\tilde{\Iulk}$.
Here, for any $\Gulc\in\tilde{\Iulk}$ we define
 $\tilde{f}^{\rm wsr}_m(\Gulc\cap\Omegaul^{(m)}) = \hat{O}(1,{\bf 1})$, where $\hat{O}(1,{\bf 1})$ is computed as described in Section \ref{sec:charwsr} for the macro TP $m$ and its set of pico TPs $\Bc_m$ using unit budgets and the given association in $\Gulc\cap\Omegaul^{(m)}$, with the following caveat.
In particular, now in obtaining the user sets
 $\{\Uc^{(b)} \} $ we treat the user in each tuple $(u,b)\in \Gulc\cap\Omegaul^{(m)}$ as a distinct virtual user. Hence, if $(u,b_1)$ and $(u,b_2)$ belong to $\Gulc\cap\Omegaul^{(m)}$, we suppose that two distinct virtual users with their own separate peak rates and associated maximum and minimum rate limits are specified. These peak rates and limits are of course identical, respectively,   to those of user $u$ and
 we
 have $\tilde{f}^{\rm wsr}(\Gulc)=f^{\rm wsr}(\Gulc),\;\forall\;\Gulc\in\Julk$. Notice that under the  admission control assumtion each member of $\Iulk$ is feasible so that $\Julk=\Iulk$. Further,   each member in $\tilde{\Iulk}$ is also feasible.   Then, we can verify from the arguments used to prove Theorem \ref{thm:wsrsubmod} that $\tilde{f}^{\rm wsr}(.)$ is a normalized non-negative submodular set function over
 $\tilde{\Iulk}$.
With this understanding, we can  re-formulate (\ref{eq:Bigoriginalwsr2}) as the following constrained set function maximization, 
 \begin{eqnarray}\label{eq:newsubmax}
 \max_{  \Gulc\in\Iulk}\{ \tilde{f}^{\rm wsr}(\Gulc)\}
 \end{eqnarray}
Then, let $\hat{\Gulc}$ be the set returned by Algorithm \ref{algo:gels} and let $\hat{\Oulc}$ denote any optimal solution of (\ref{eq:newsubmax}). Notice that $\hat{\Gulc}\cup\hat{\Oulc}\in\tilde{\Iulc}$ so that the extended set function is defined and is submodular over all subsets of  $\hat{\Gulc}\cup\hat{\Oulc}$.
  This enables us to invoke the arguments presented in \cite{Lee:submod}  to prove the aproximation guarantee  and thereby establish our desired result.\endproof
\begin{corollary}
In the absence of per-user minimum rate constraints the problem in (\ref{eq:Bigoriginalwsr2}) reduces to the monotone submodular maximization problem with a matroid constraint, for which the greedy stage output itself (and thus Algorithm \ref{algo:gels}) yields an approximation factor of $1/2$.
\end{corollary}
Regarding the worst-case complexity of Algorithm \ref{algo:gels}, we can show that it scales polynomially in $|\Omega|/\epsilon$ \cite{Lee:submod} . In practice, in all of our simulation runs we observed that with the greedy stage initialization the enhanced LS stage converges very quickly. Moreover, the implementation of the greedy stage of Algorithm \ref{algo:gels} can be significantly improved by exploiting the submodularity of $f^{\rm wsr}(.)$ over $\Julc$, as done in the {\em lazy greedy} implementation \cite{Min:lazy}.

Let us now focus on the  problem in (\ref{eq:Bigoriginal}). In order to design an approximation algorithm, we
consider the problem in (\ref{eq:Bigoriginal2}), where we recall our convention that
$0\ln(0)=0$. Note that (\ref{eq:Bigoriginal2}) imposes an orthogonal split on (\ref{eq:Bigoriginal}) and allows for each user to be associated to (and served by) exactly one node. The problem in (\ref{eq:Bigoriginal2})  has been widely considered before   and seeks to optimize the PF utility over user associations but does not permit dual connectivity.  
There are several approaches to solve (\ref{eq:Bigoriginal2}),  including an efficient optimal one \cite{prasad:lb}  and approximately optimal ones with lower complexity \cite{shen:lb,ye:ua,prasad:lb}. In Algorithm \ref{algo:osp} we propose a method to solve (\ref{eq:Bigoriginal}) where we can leverage any of  the available approaches to solve (\ref{eq:Bigoriginal2}). Once a user association is so obtained, we enhance it by exploiting dual connectivity. Hence, all users associated  to any pico node $b\in\Bc_m$ for any $m\in\Mc$ are also   connected to the macro $m$. Further,  each user associated to a macro TP is also associated to a  pico TP in the set of pico TPs assigned to that macro. Then, the allocation fractions are optimized as described in  Section \ref{sec:charpf}. The performance guarantee of Algorithm \ref{algo:osp} is established below,  where we let
  $\Pi\geq 0$ denote the (additive) guarantee pertaining to the approach used to solve  (\ref{eq:Bigoriginal2}), i.e., the value yielded by the obtained output is no less than the corresponding  optimal objective value of (\ref{eq:Bigoriginal2}) minus $\Pi$ (so that $\Pi=0$ for the optimal algorithm \cite{prasad:lb}). The key insight used in the proof is from Corollary \ref{coro:pf} that for any valid association, considering each macro and each pico assigned to that macro,  optimal allocation fractions entail only one user receiving a positive resource share from both that macro and  pico.
\begin{figure*}{{\begin{equation}\boxed{\begin{aligned}\label{eq:Bigoriginal2}
   \max_{x_{u,b}\in\{0,1\}\;\forall\;u\in\Uc,b\in\Sc}\left\{\sum_{u\in\Uc}\sum_{b\in\Sc}x_{u,b}\ln\left( R_{u,b}\right)- \sum_{b\in\Sc}\left(\sum_{k\in\Uc}x_{k,b}\right)\ln\left(\sum_{k\in\Uc}x_{k,b}\right)     \right\}\;\;
 {\rm s.t. } \sum_{b\in\Sc}x_{u,b}= 1,\;\forall\;u\in\Uc. \\
\end{aligned}}\end{equation}}\vspace{-1cm}}        \end{figure*}

\begin{theorem}\label{prop:bound}
\vspace{-.5cm} Algorithm \ref{algo:osp} provides an output that     yields  an objective value for (\ref{eq:Bigoriginal}) that is no less than the optimal objective value of (\ref{eq:Bigoriginal}) minus $\Pi + \min\{K,\sum_{m\in\Mc}|\Bc_m|\}\ln(2)$.
\end{theorem}
\proof  We first note that (\ref{eq:Bigoriginal2})  is equivalent to (\ref{eq:Bigoriginal}) with the additional constraint that $\gamma_{u,b}\theta_{u,m}=0,\;\forall\;u\in\Uc,b\in\Bc_m,m\in\Mc$. Suppose that (\ref{eq:Bigoriginal2}) is solved using an approach that offers a guarantee of $\Pi$.\footnote{Note that since the objective function in (\ref{eq:Bigoriginal2}) and (\ref{eq:Bigoriginal}) can be negative, we can only offer additive guarantees instead of multiplicative ones.}  Then, note that the obtained solution  feasible for (\ref{eq:Bigoriginal}) and invoking  Proposition \ref{prop:solnsimpO} (once for each macro TP together with its assigned set of pico TPs and the users associated to them) we can conclude that the attained objective value is no less than the optimal one minus the claimed additive factor. The remaining steps of Algorithm \ref{algo:osp} further improve the solution at hand and hence further reduce the gap to optimal, which proves the theorem. \endproof
We remark that another way to view the performance guarantee of Algorithm \ref{algo:osp} (when $\Pi=0$) is as follows. Let us scale all the peak rates of any set of $\min\{K,\sum_{m\in\Mc}|\Bc_m|\}$ users by $2$ and obtain an output  by Algorithm \ref{algo:osp}. Then, the objective value in (\ref{eq:Bigoriginal}) yielded by the solution at hand, will be no less than the one yielded by the optimal solution using the original peak rates.

\begin{table}
\caption{{\bf GELS Algorithm}}\label{algo:gels}
\begin{small}
\begin{algorithmic}[1]
\STATE Initialize with  ${\rm MaxIter}\geq 1$, $\tilde{\Omegaul}=\Omegaul$, $\Delta>0$ and  $\hat{\Gulc}=\phi$.
\STATE \textbf{Repeat} $\;\;\;\%{\rm Optional\;\;Greedy \;\;Stage}$
\STATE Determine $(k',b')$ as the tuple in $\tilde{\Omegaul}\setminus\hat{\Gulc}$ which satisfies $\hat{\Gulc}\cup(k',b')\in\Julk$ and
 offers the best gain $f^{\rm wsr}(\hat{\Gulc}\cup(k',b')) - f^{\rm wsr}(\hat{\Gulc})$, among all such tuples. 
\STATE If the best gain is positive update $\hat{\Gulc}=\hat{\Gulc}\cup(k',b')$.
 \STATE   \textbf{Until} the best gain is not positive. 
\STATE Set $\breve{\Gulc}=\hat{\Gulc}$, ${\rm Iter=0}$.
\STATE \textbf{Repeat} $\;\;\%{\rm Enhanced \;\;Local \;\;Search \;\;(LS) \;\;Stage}$
\STATE Increment ${\rm Iter}={\rm Iter}+1$.
\STATE {\em Swap:} Find a pair of tuples: $(k',b')\in\breve{\Gulc}$ and $(k,b)\in\tilde{\Omegaul}\setminus\breve{\Gulc}$ such  that the  swapping $(k',b')\in\breve{\Gulc}$ with $(k,b)$ is in $\Julk$ and results in the best gain among such swaps.
\STATE {\em Deletion:} Find a  tuple: $(k',b')\in\breve{\Gulc}$ such that deleting this tuple results in the best gain among such deletions.
\STATE {\em Addition:} Find a  tuple: $(k',b')\in\tilde{\Omegaul}\setminus\breve{\Gulc}$ such that $\breve{\Gulc}\cup(k',b')\in\Julk$ and adding this tuple results in the best gain among such additions.
\STATE Determine the overall best among the three best gains and compute threshold $\Delta f^{\rm wsr}(\breve{\Gulc}) $
   \STATE If  the overall best  gain is at-least as as large as the threshold, then
 update $\breve{\Gulc}$ according to the overall best choice.
 \STATE   \textbf{Until} the overall  best gain is less than threshold or ${\rm Iter}={\rm MaxIter}$.
\STATE Repeat Greedy and Enhanced LS stages intialized with $\tilde{\Omegaul}=\Omegaul\setminus\breve{\Gulc}$ and let $\tilde{\Gulc}$ be the obtained output.
\STATE Output  the choice yielding the larger of $\{f^{\rm wsr}(\breve{\Gulc}),f^{\rm wsr}(\tilde{\Gulc})\}$.
\end{algorithmic}
\end{small}\vspace{-.5cm}
\end{table}

\begin{table}
\caption{{\bf  Orthogonal Split Processing based Algorithm (OSPA)}}\label{algo:osp}
\begin{algorithmic}[1]
\STATE Initialize with  $\Uc,\Mc,\Bc_m,\forall\;m\in\Mc$.
\STATE Set $\Sc=\Mc\cup(\cup_{m\in\Mc}\Bc_m)$ and determine user associations $\{x_{u,b}\},u\in\Uc,b\in\Sc$ by solving (\ref{eq:Bigoriginal2})
\STATE \textbf{For} each macro TP $m\in\Mc$ \textbf{Do}
\STATE Consider each user with $x_{u,m}=1$ and  associate that user with the pico TP in $\Bc_m$ yielding the strongest received power for that user.
\STATE  Using the obtained association for TPs in $\Bc_m$ obtain the optimal allocation fractions using Theorem \ref{thm:pfalloc}.
\STATE \textbf{End For}
\STATE Output  the  user associations and allocation fractions.
\end{algorithmic}  \vspace{-.5cm}
\end{table}

\section{Simulation Results}

We now present our   simulation results obtained for an LTE HetNet deployment. We emulate a HetNet comprising of $57$ macro cells with $10$ pico cells  being assigned to each macro and with a full buffer traffic model. 
Each macro  base-station transmits with a power of $46\;dBm$ whereas the transmit power at each pico node is $40\;dBm$ and the system bandwidth is 10 MHz. A noise PSD of $-174\;dB/Hz$ with a noise figure of $9\;dB$ were assumed.  Other major parameters such as the distributions used to drop users, macro and pico nodes  are all as per 3GPP guidelines. 

In the first set of results we consider the weighted sum rate system utility optimization (\ref{eq:Bigoriginalwsr2}) and set all user weights to be unity. We use the network setting described above and focus on the out-of-band low load scenario (342 users). As will shortly be demonstrated the gains of DC are more pronounced for this choice. 
We suppose that there is no upper bound on the rate for any user so that the per-user maximum rate constraints are all vacuous. We set the minimum rates so that the admission control assumption is satisfied. We compare our proposed  GELS algorithm  with a baseline single point association scheme in which each  user independently associates to the TP from which it can obtain the highest peak rate. This association scheme is also referred to as the maximum SINR association   \cite{andrews:LBmag}. Furthermore, in this baseline scheme each TP adopts a round robin policy to serve its associated users. Notice that we cannot enforce any minimum rates on this baseline scheme. On the other hand, we implement our algorithm on three different cluster sizes: (i) each cluster of size 11 including  one macro along with 10 pico TPs assigned to that macro, (ii) each clutser of size 33 comprising of three macros and 10 picos assigned to each macro, respectively, and (iii) one cluster of size 627 comprising of 57 macros and 10 picos assigned to each macro in that cluster. In Fig. \ref{fig:DCclusters} we plot the average cell spectral efficiency (SE)  per macro cell for all the three different cluster sizes. 
The key takeaway is that compared to the baseline, DC offers a large improvement and that most of this improvement is captured by a small cluster size. 
\begin{figure}[!tbp]
  \centering
  \begin{minipage}[b]{0.45\textwidth}
    \includegraphics[width=\textwidth]{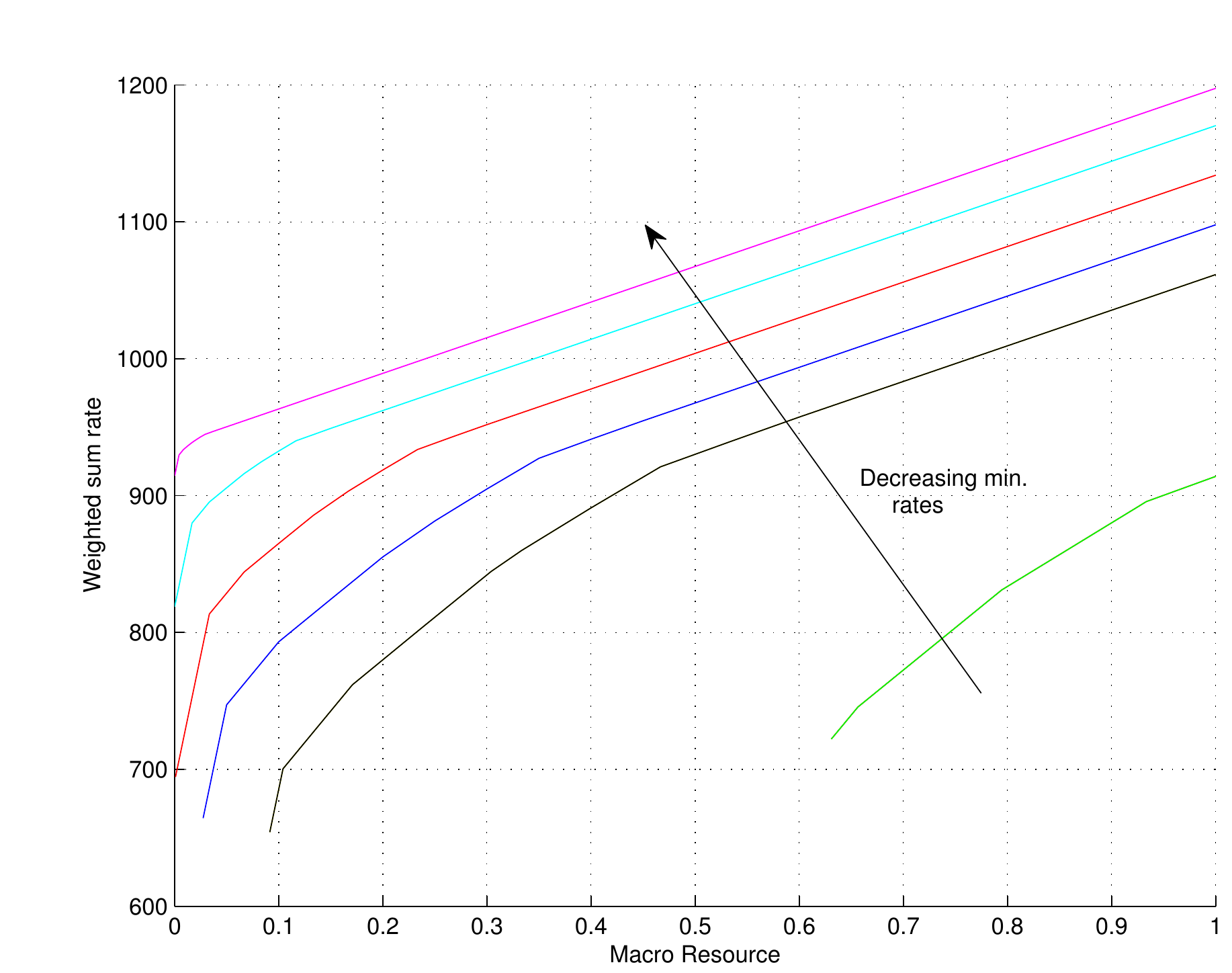}
    \caption{Optimized weighted sum rate vs macro budget for different min. rates}\label{fig:DCwsr}
  \end{minipage}
  \hfill
  \begin{minipage}[b]{0.45\textwidth}
    \includegraphics[clip, trim=3cm 5cm 2cm 2cm,width=0.7\textwidth]{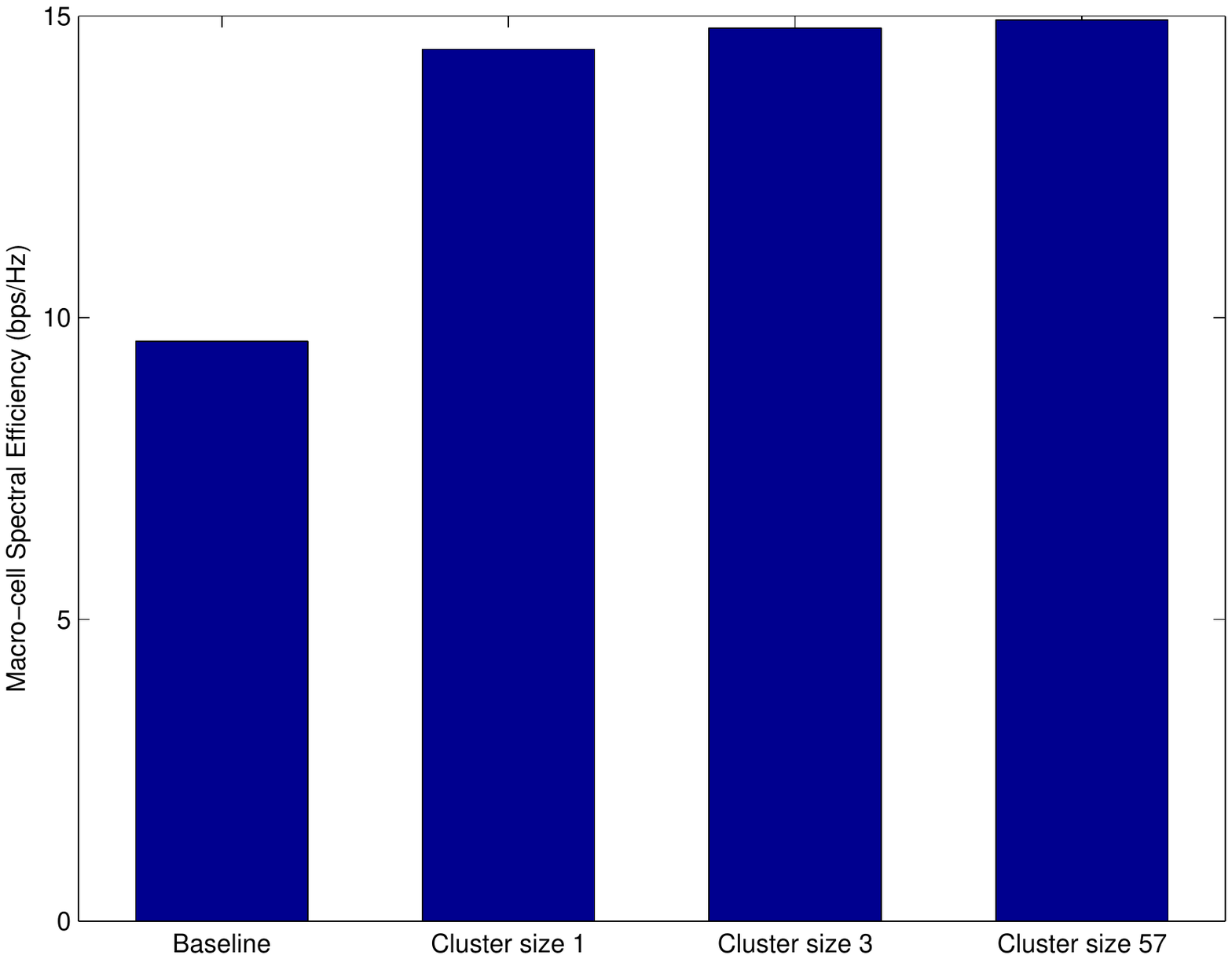}
    \caption{Cell SE for different cluster sizes}\label{fig:DCclusters}
  \end{minipage} 
\end{figure}
We next report the performance of  OSPA which optimizes the PF utility  (\ref{eq:Bigoriginal}) and consider an in-band scenario as well as an out-of-band scenario.
 To benchmark the performance of this algorithm,   we  determine the average  and the $5-$percentile spectral efficiency (SE) yielded by the single-point max-SINR baseline   scheme.   
Next, we determine the   average and $5-$percentile SE values yielded by the   user association (UA) algorithm from \cite{prasad:lb} that optimizes the PF utility without exploiting DC (\ref{eq:Bigoriginal2}). Finally, we use that algorithm as a module in OSPA to optimize (\ref{eq:Bigoriginal2}), with the obtained output being further refined   by exploiting DC. 
 The obtained results    are plotted in Figs. \ref{fig:DCIn} and \ref{fig:DCOut}  as relative percentage gains over the respective baseline counterparts, for the in-band and out-of-band scenarios, respectively. In each figure we consider three different load points, such the first load point emulates a HetNet with 342 users, the second one has 684 users and  the last load point has  1368 users, respectively.
From the results in  these figures, we see that  DC can be quite beneficial at low to moderate loads, which is intuitively satisfying.


\section{Conclusions}
We considered the problem of maximizing the weighted sum rate and the proportional fairness utility over  dual connectivity enabled HetNets  by   exploiting  load balancing. We  constructed efficient algorithms to solve the resulting mixed optimization problems and proved that they yield approximately optimal solutions. 

\appendix 
\begin{definition}
  $(\Omega,\Iul)$, where $\Omega$ is a ground set and $\Iul$ is collection of some subsets of $\Omega$, is said to be a {\em matroid} if 
  \begin{itemize}
 \item $\Iul$ is downward closed, i.e., $\Ac\in\Iul\;\&\;\Bc\subseteq\Ac\Rightarrow\Bc\in\Iul$
  \item For any two members $\Fc_1\in\Iul$ and $\Fc_2\in\Iul$ such that $|\Fc_1|<|\Fc_2|$, there exists
   $e\in\Fc_2\setminus\Fc_1$ such that $\Fc_1\cup\{e\}\in\Iul$. This property is referred to as the exchange property.
 \end{itemize}
\end{definition}
\begin{definition}
Let $\Jul$ be a collection of some subsets of $\Omega$ and  $h:\Jul\to\Reals$  be a real-valued set function defined on the members of $\Jul$. The set function
$h(.)$ is   a {\em submodular} set function over $\Jul$ if   it satisfies,
$h(\Bc\cup a)-h(\Bc)\leq h(\Ac\cup a)-h(\Ac)$ 
for all $\Ac\subseteq\Bc\in\Jul\;\& \;a\in\Omega\setminus\Bc:\Bc\cup a\in\Jul.$
\end{definition}
The following lemma is invoked in the proof of Proposition \ref{prop:wsrSs} and its proof 
  follows from the arguments used in that of Proposition \ref{prop:charsub}. 
\begin{lemma}\label{lem:rel}
For any TP $b\in\Bc_1$ with any budget $\Gamma_b\leq 1$, the slope curve  $S(Z_b,\Gamma_b,b) $ is piecewise constant and decreasing in $Z_b$ for all $Z_b\geq \bar{h}(\Gamma_b,b)$. Further, for any given 
 $\Gamma_b$ and any scalar $\delta_b\leq \Gamma_b$, we have 
\begin{eqnarray}\label{eq:slopred}
S(Z_b,\Gamma_b,b)\leq S(Z_b,\Gamma_b-\delta_b,b),\;\forall\;Z_b\geq \bar{h}(\Gamma_b-\delta_b,b)
\end{eqnarray}
 For any given $Z_b$, let   $T(Z_b,\Gamma_b,b) $ denote the slope (with respect to the pico resource) of the  function $\hat{O}(Z_b,\Gamma_b,b)$ at any $\Gamma_b:\Gamma_b\geq \hul(Z_b,b)$. In particular,
\begin{eqnarray}\label{eq:wsrspP}
T(Z_b,\Gamma_b,b) = \lim_{\delta\to 0_+}\frac{\hat{O}(Z_b,\Gamma_b+\delta,b)- \hat{O}(Z_b,\Gamma_b,b)}{\delta}
\end{eqnarray}
Then, for any $Z_b\leq 1$, the slope curve $ T(Z_b,\Gamma_b,b)$ is piecewise constant and decreasing in $\Gamma_b$ for all
 $\Gamma_b\geq \hul(Z_b,b)$. Further, for any $z_b\leq Z_b$, we have that  
\begin{eqnarray}\label{eq:slopredP}
T(Z_b,\Gamma_b,b)\leq T(Z_b-z_b,\Gamma_b,b),\;\forall\;\Gamma_b\geq \hul(Z_b-z_b,b). 
\end{eqnarray}
\end{lemma}
{\textbf{Proof of Proposition \ref{prop:wsrSs}}}: 
 We will first prove the claim for $b_1\neq b_2$ and then  consider the case $b_1=b_2$.
Consider the LHS of (\ref{eq:subnomax}) and parse the difference as
\begin{eqnarray*}
  \hat{O}(\Gamma,\Gammab) - \hat{O}(\Gamma+\delta,\Gammab+\delta_{b_1}\eb_{b_1}) = \underbrace{\hat{O}(\Gamma,\Gammab) - \hat{O}(\Gamma,\Gammab+\delta_{b_1}\eb_{b_1})}
 + \underbrace{\hat{O}(\Gamma,\Gammab+\delta_{b_1}\eb_{b_1}) -  \hat{O}(\Gamma+\delta,\Gammab+\delta_{b_1}\eb_{b_1})   }
\end{eqnarray*}
The RHS of (\ref{eq:subnomax}) can be parsed analogously. Then, it suffices to show that
\begin{eqnarray}\label{eq:subnomaxB}
 \hat{O}(\Gamma,\Gammab) - \hat{O}(\Gamma,\Gammab+\delta_{b_1}\eb_{b_1}) \leq
  \hat{O}(\Gamma+\tilde{\delta},\Gammab+ \tilde{\delta}_{b_2}\eb_{b_2}) - \hat{O}(\Gamma+\tilde{\delta},\Gammab+ \tilde{\delta}_{b_2}\eb_{b_2}+\delta_{b_1}\eb_{b_1})
\end{eqnarray}
and that
\begin{eqnarray}\label{eq:subnomaxC}
 \nonumber  \hat{O}(\Gamma,\Gammab+\delta_{b_1}\eb_{b_1}) -  \hat{O}(\Gamma+\delta,\Gammab+\delta_{b_1}\eb_{b_1})\;\;\;\;\;\;\;\;\;\;\;\;\;\;\;\;\;\;\;\;\;\;\;\;\;\;\;\;\;\;\;\;\;\;\;\;\;\;\;\qquad\qquad\qquad\qquad\\ \leq
 \hat{O}(\Gamma+\tilde{\delta},\Gammab+\tilde{\delta}_{b_2}\eb_{b_2}+\delta_{b_1}\eb_{b_1}) -  \hat{O}(\Gamma+\tilde{\delta}+\delta,\Gammab+ \tilde{\delta}_{b_2}\eb_{b_2} + \delta_{b_1}\eb_{b_1})
\end{eqnarray}
We first consider (\ref{eq:subnomaxC}). Letting $\tilde{\Gamma}=\Gamma+\tilde{\delta}, \tilde{\Gammab}=\Gammab+\delta_{b_1}\eb_{b_1}$,  $\breve{\Gammab}=\tilde{\Gammab}+\tilde{\delta}_{b_2}\eb_{b_2}=\Gammab+\delta_{b_1}\eb_{b_1}+\tilde{\delta}_{b_2}\eb_{b_2}$,
 suppose that the following relation holds.
\begin{eqnarray}\label{eq:sloperel}
S(Z,\breve{\Gammab}) \leq S(Z,\tilde{\Gammab}),\;\forall\; Z\geq \sum_{b\in\Bc_1'}\bar{h}(\tilde{\Gamma}_b,b).
\end{eqnarray}
Then, invoking (\ref{eq:corosubmain}) the relation in (\ref{eq:subnomaxC}) can also be expressed as
\begin{eqnarray}\label{eq:lhsint}
 \int^{\tilde{\Gamma}+\delta}_{\tilde{\Gamma}} S(z,\breve{\Gammab} )dz \leq
 \int^{\Gamma+\delta}_{\Gamma} S(z,\tilde{\Gammab}  )dz.
\end{eqnarray}
The relation in (\ref{eq:lhsint}) indeed holds as a result of (\ref{eq:sloperel}) and the fact that $\Gamma\leq \tilde{\Gamma}=\Gamma+\tilde{\delta}$ and that the slope curve $S(z,\tilde{\Gammab})$ is non-increasing in $z$. Notice that the integral in the LHS of (\ref{eq:lhsint}) is the area covered under rectangles  of non-increasing heights $S(z,\breve{\Gammab}), z\in [\tilde{\Gamma}, \tilde{\Gamma}+\delta]$ and total width $\delta$. An analogous observation holds for the
  RHS of (\ref{eq:lhsint}).
It remains to prove for this case that (\ref{eq:sloperel}) indeed is true. We accomplish this via contradiction.
 and suppose that $\exists Z'  \geq \sum_{b\in\Bc_1'}\bar{h}(\tilde{\Gamma}_b,b):
S(Z',\breve{\Gammab}) > S(Z',\tilde{\Gammab})$. Then, notice that
$\sum_{b\in\Bc_1'}\bar{h}(\tilde{\Gamma}_b,b)\geq \sum_{b\in\Bc_1'}\bar{h}(\breve{\Gamma}_b,b)$.
Consider next how macro resources are assigned by Algorithm \ref{algo:wsrallocopt} for given pico budgets $\breve{\Gammab},\tilde{\Gammab}$. We can see that for both pico budget vectors,
 the macro resources to all TPs in $\Bc_1'\setminus b_2$ are assigned as slack in the same order, i.e.,  if a TP $b\neq b_2$ is assigned a macro resource share for the $n^{th}$ time (for any $n\geq 1$) under $\breve{\Gammab}$ then that   TP can be  assigned a macro resource share for the $n^{th}$ time under $\tilde{\Gammab}$ only after all prior macro assignments made to TPs in $\Bc_1'\setminus b_2$ under $\breve{\Gammab}$ have been made to those TPs  under $\tilde{\Gammab}$ as well.  Then, if upto $Z'$\\
$\bullet$ At-least as much macro resource has been assigned as slack to TPs in $\Bc_1'\setminus b_2$ under $\breve{\Gammab}$ than under $\tilde{\Gammab}$, we immediately have the contradiction since $S(Z',\tilde{\Gammab})$ must be at least as large as the slope value at the most recent assignment to made to any TP in  $\Bc_1'\setminus b_2$ under $\breve{\Gammab}$. The latter slope value in turn must be no less than $S(Z',\breve{\Gammab})$, since $S(Z,\breve{\Gammab})$ is non-increasing in $Z$.\\
$\bullet$  Less macro resource has been assigned as slack to TPs in $\Bc_1'\setminus b_2$ under $\breve{\Gammab}$ than under $\tilde{\Gammab}$, so that at-least $\bar{h}(\tilde{\Gamma}_{b_2},b_2) - \bar{h}(\breve{\Gamma}_{b_2},b_2)$ more macro resource as slack has been assigned as slack to TP  $b_2$ under $\breve{\Gammab}$. Invoking (\ref{eq:slopred}) (for $b=b_2$) we again immediately have the contradiction since $S(Z',\tilde{\Gammab})$ must be at least as large as the slope value at the most recent assignment to made to   TP in  $ b_2$ under  $\breve{\Gammab}$. 
  The latter slope value in turn must be no less than $S(Z',\breve{\Gammab})$  since $S(Z,\breve{\Gammab})$ is non-increasing in $Z$.\\

Let us now proceed to prove (\ref{eq:subnomaxB}).
Towards this end, let $\Delta_{b_2}=\sum_{b\in\Bc_1'}\bar{h}(\tilde{\Gamma}_b,b) - \sum_{b\in\Bc_1'}\bar{h}(\breve{\Gamma}_b,b)$ and let $\Psi_{b_2}$ denote the total macro resource assigned as slack
 to TP $b_2$ under $\breve{\Gammab}$.   Following Algorithm \ref{algo:wsrallocopt}
 we note that  $S(Z,\tilde{\Gammab}),\;\forall\; Z\in \left[\sum_{b\in\Bc_1'}\bar{h}(\tilde{\Gamma}_b,b),\tilde{\Gamma}+\delta\right]$ can be obtained from
$S(Z,\breve{\Gammab}),\;\forall\; Z\in \left[\sum_{b\in\Bc_1'}\bar{h}(\breve{\Gamma}_b,b),\tilde{\Gamma}+\delta\right]$ by 
 performing the following  steps:\\
$\bullet$ Expurgate    Macro  allocations to TP $b_2$   without leaving any holes (i.e., an allocation to TP $b_2$  is expurgated only after all preceding   allocations to TP $b_2$ have been expurgated) such that the width of those expurgated allocations equals $\min\{\Psi_{b_2},\Delta_{b_2}\}$.\\
$\bullet$ If  $\Psi_{b_2}< \Delta_{b_2}$, then
 expurgate the allocations corresponding to $Z\in [\tilde{\Gamma}+\delta-(\Delta_{b_2} - \Psi_{b_2}),\tilde{\Gamma}+\delta]$.\\
$\bullet$ If  $\Psi_{b_2}\geq  \Delta_{b_2}$, then
 replace slope values of all remaining allocations  to TP $b_2$  (with width $\Psi_{b_2} -  \Delta_{b_2}$) using values (in order) from $S(Z_{b_2},\tilde{\Gamma}_{b_2},b_2),Z_{b_2}\geq \bar{h}(\tilde{\Gamma}_{b_2},b_2)$.\\
$\bullet$ Using values (in order) from $S(Z_{b_2},\tilde{\Gamma}_{b_2},b_2),Z_{b_2}\geq \bar{h}(\tilde{\Gamma}_{b_2},b_2) + (\Psi_{b_2}-\Delta_{b_2})^+$ replace lower slope values (if possible) corresponding to allocations other than TP $b_2$, where $(x)^+=\max\{0,x\},\;\forall\;x\in\Reals$.\\
Note that the expurgated allocations in the second step (if any) necessarily do not belong to TP $b_2$. Together, the area covered by the expurgated slope values in the first and second step represent a loss. Further, invoking (\ref{eq:slopred}) (for $b=b_2$) proved in Lemma \ref{lem:rel} we see that the replacing of slope values  if done in the third step certainly results in an improvement. On the other hand, by construction the replacing of slope values  if done in the fourth step  results in an improvement.

Let us apply the same procedure to obtain $S(Z,\Gammab),\;\forall\; Z\in \left[\sum_{b\in\Bc_1'}\bar{h}(\Gamma_b,b),\Gamma \right]$   from
$S(Z,\tilde{\Gammab}),\;\forall\; Z\in \left[\sum_{b\in\Bc_1'}\bar{h}(\tilde{\Gamma}_b,b),\Gamma\right]$
and to
obtain $S(Z,\hat{\Gammab}),\;\forall\; Z\in \left[\sum_{b\in\Bc_1'}\bar{h}(\hat{\Gamma}_b,b),\tilde{\Gamma}\right]$    from
$S(Z,\breve{\Gammab}),\;\forall\; Z\in \left[\sum_{b\in\Bc_1'}\bar{h}(\breve{\Gamma}_b,b),\tilde{\Gamma}\right]$, where we define $\hat{\Gammab}= \Gammab + \delta_{b_2}\eb_{b_2}$.
 In either case the net loss
  is equal to $H(\tilde{\Gamma}_{b_1},b_1)-H(\Gamma_{b_1},b_1)$ plus the loss of the first two steps and minus the improvement of the last two steps.
 Now,  to see that
 the relation in (\ref{eq:subnomaxB}) holds, we can ignore the common term $H(\tilde{\Gamma}_{b_1},b_1)-H(\Gamma_{b_1},b_1)$ and first verify that
 the loss (or area under the expurgated slope values) when   obtaining  $S(Z,\Gammab),\;\forall\; Z\in \left[\sum_{b\in\Bc_1'}\bar{h}(\Gamma_b,b),\Gamma \right]$   from
$S(Z,\tilde{\Gammab}),\;\forall\; Z\in \left[\sum_{b\in\Bc_1'}\bar{h}(\tilde{\Gamma}_b,b),\Gamma\right]$ is at least as large as the loss on obtaining $S(Z,\hat{\Gammab}),\;\forall\; Z\in \left[\sum_{b\in\Bc_1'}\bar{h}(\hat{\Gamma}_b,b),\tilde{\Gamma} \right]$   from
$S(Z,\breve{\Gammab}),\;\forall\; Z\in \left[\sum_{b\in\Bc_1'}\bar{h}(\breve{\Gamma}_b,b),\tilde{\Gamma}\right]$. This is because the expurgated slope values in the former case are at least as large,  whereas the  width (or the macro resource spanned under the expurgated slopes) is identical in each case to $\bar{h}(\Gamma_{b_1},b_1) - \bar{h}(\Gamma_{b_1}+\delta_{b_1},b_1)$.  On the other hand,
the improvement obtained in the former case is no greater than the latter case. This follows from the facts that while the same set of slope values $S(Z_{b_1},\Gamma_{b_1},b_1),\;\forall\; Z_{b_1}\geq  \bar{h}(\Gamma_{b_1},b_1)$ are used to replace existing ones in each case, in the former case the replaced slope values  are at least as large while the width (or the macro resource spanned under the replaced slopes) is   no greater, 
thereby resulting in a  smaller improvement. Taking these observations together we have the  desired result in
 (\ref{eq:subnomaxB}).

Let us now consider the case $b_1=b_2$.  
The method we employ is to reformulate the problem at hand into one involving only one pico TP ($b_1$), albeit with an expanded pool of associated users, and the macro TP. Towards this end, define a set of pico TPs $\hat{\Bc}_1=\Bc'_1\setminus b_1$.  Then, specializing the result in (\ref{eq:corosubmain})   to the set of pico TPs $\hat{\Bc}_1$  along with the set of users associated to those TPs, we notice that the weighted sum rate is the sum of three terms. The first two terms are the weighted sums of  minimum rates that have to be met and the rates obtained upon assigning the pico  resource as slack (if any), respectively.  The third term pertains to the weighted sum rate obtained by assigning macro resource as slack. As the first two terms are invariant to changes in $\Gamma_{b_1}$ we focus on the third term. Note that the latter term is the area under rectangles of decreasing heights, where these heights are the distinct slope values of the curve $\hat{O}(Z,\{\Gamma_{b}\}_{b\in\hat{\Bc}_1}), Z\in [\sum_{b\in\hat{\Bc}_1}\bar{h}(\Gamma_b,b),1]$. For each such rectangle let us define a virtual user, say $\vartheta$ and set the parameters of that virtual user as follows.  Arbitrarily choose its weight, $w_\vartheta$, and peak rate from the macro TP, $R_{\vartheta,1}$, to be any positive scalars such that  $w_\vartheta R_{\vartheta,1}$ equals the height of that rectangle and set $R^{\rm max}_\vartheta$ to be equal to  $R_{\vartheta,1}$ times the width of the rectangle. Further, set $R_\vartheta^{\rm min}=0,R_{\vartheta,b_1}=0$. Collecting all such virtual users in a set $\hat{\Uc}$, consider the problem of allocating the macro resource when we have just one pico TP $b_1$ but to which an expanded user pool $\Uc^{(b_1)}\cup\hat{\Uc}$ is associated. By our construction of the virtual user set 
 it can be readily seen that for any feasible pico budgets, this reformulated problem is equivalent to the original one (\ref{eq:Bigoriginalwsrp}).
Moreover, in this reformulated problem the pico resource of TP $b_1$ is assigned to only users in $\Uc^{(b_1)}$. We can invoke Proposition \ref{prop:charsub} on this reformulated problem along with the relations (\ref{eq:slopred}) and (\ref{eq:slopredP}) to deduce that the claims in 
(\ref{eq:subnomaxC}) and (\ref{eq:subnomaxB}) indeed hold true. \qed
 
{\textbf{Proof of Theorem \ref{thm:pfalloc}}}: 
 The problem in (\ref{eq:Bigoriginalp2}) is a convex optimization problem for which the K.K.T conditions are both necessary and sufficient.
These K.K.T conditions include: 
\begin{eqnarray}\label{eq:condkkt}
  \frac{R_{u,1}}{R_{u,1}\theta_{u,1}+R_{u,b}\gamma_{u,b}}+\eta_{u,1}=\lambda;\;
 \frac{R_{u,b}}{R_{u,1}\theta_{u,1}+R_{u,b}\gamma_{u,b}}+\vartheta_{u,b}=\beta_b, \;\forall\;u\in\Uc^{(b)},b\in\Bc_1
   \end{eqnarray}
 along with the
complementary slackness conditions
 $\eta_{u,1}\theta_{u,1}=0,\vartheta_{u,b}\gamma_{u,b}=0,\;\forall\;u\in\Uc^{(b)},b\in\Bc_1'$ and
$ (1-\sum_{u\in\Uc'}\theta_{u,1})\lambda=0$, $(1-\sum_{u\in\Uc^{(b)}}\gamma_{u,b})\beta_b=0,\;\forall\;b\in\Bc_1'$, and the  feasibility ones  $\theta_{u,1},\gamma_{u,b}\in [0,1],\;\forall\;u,b$,    $\sum_{u\in\Uc'}\theta_{u,1}\leq 1,\sum_{u\in\Uc^{(b)}}\gamma_{u,b}\leq 1,\forall\;b$. Note that
 $\lambda,\{\beta_b\},\{\eta_{u,1}\},\{\vartheta_{u,b}\}$ are non-negative Lagrangian variables. Manipulating the first two K.K.T conditions in (\ref{eq:condkkt}) along with the complementary slackness conditions, we see that for any two distinct users $u,u'\in\Uc^{(b)}$
 \begin{eqnarray*}\label{eq:condkkt2}
 \theta_{u,1}\gamma_{u,b}> 0\;\;\&\;\;\theta_{u',1}\gamma_{u',b}> 0\Rightarrow \frac{R_{u,1}}{R_{u,b}}=\frac{R_{u',1}}{R_{u',b}},
 \end{eqnarray*}
which is a contradiction. Consequently, there can be at-most one user $u\in\Uc^{(b)}$ in each $b\in\Bc_1'$ for which $\theta_{u,1}\gamma_{u,b}> 0$.
 From the remaining conditions, we can further deduce that
\begin{eqnarray*}\label{eq:condkkt2}
 \theta_{u,1}=0\Rightarrow \gamma_{u,b}=\frac{1}{\beta_b}\;\;\&\;\;     \frac{R_{u,1}}{R_{u,b}}\leq  \frac{\lambda}{\beta_b};\;\;
 \gamma_{u,b}=0\Rightarrow \theta_{u,1}=\frac{1}{\lambda}\;\;\&\;\;    \frac{R_{u,1}}{R_{u,b}}   \geq  \frac{\lambda}{\beta_b}
   \end{eqnarray*}
Further, an optimal solution must fully use the available resource  so that $\sum_{u\in\Uc'}\theta_{u,1}=1$
 and $\sum_{u\in\Uc^{(b)}}\gamma_{u,b}=1,\;\forall\;b\in\Bc_1'$.
Next, for each $b\in\Bc_1'$, let us determine the
scalars $\mu_{m,b},m=1,\cdots,N_b$  by sorting the set of ratios $\{ \frac{R_{u,1}}{R_{u,b}}\}_{u\in\Uc^{(b)}}$ in the increasing order. Then, define the
set $\Sulc^A_b=\cup_{m=1,\cdots,N_b}((m-1)\mu_{m,b},m\mu_{m,b})$ along with
 $\Sulc_b^B= \cup_{m=2,\cdots,N_b+1}[(m-1)\mu_{m-1,b},(m-1)\mu_{m,b}]$, which together partition the set of positive real numbers into two non-overlapping parts.
Notice that if $\lambda \in \Sulc^A_b$ then the only solution to the K.K.T conditions pertaining to TP $b$ must have
 exactly one user $k\in\Uc^{(b)}$ with $\theta_{k,1}\gamma_{k,b}> 0$ and this user must  be the one
 which has the $m^{th}$ smallest ratio $\mu_{m,b}$,  where
 $m\in\{1,\cdots,N_b\}: \lambda\in ((m-1)\mu_{m,b},m\mu_{m,b})$. Each user $u$ whose corresponding ratio $\frac{R_{u,1}}{R_{u,b}}$ is smaller than $\mu_{m,b}$ is assigned resource only in TP $b$ whereas each user $u$ whose corresponding ratio $\frac{R_{u,1}}{R_{u,b}}$ is greater than $\mu_{m,b}$ is assigned resource only by the Macro.  In this case, the total  load imposed by users associated to TP $b$ on the Macro is given by $\frac{N_b}{\lambda} - \frac{1}{\mu_{m,b}}$.
  Similarly, whenever $\lambda \in \Sulc^B_b$ then the only solution to the K.K.T consitions pertaining to TP $b$ must have
 an orthogonal split of associated users in that each user associated to TP $b$ is assigned resource either by the Macro or by TP $b$. In particular, when
 $\lambda\in [(m-1)\mu_{m-1,b},(m-1)\mu_{m,b}]$ for some $m=2,\cdots,N_b+1$, then each user $u$ whose corresponding ratio $\frac{R_{u,1}}{R_{u,b}}$ is smaller than $\mu_{m,b}$ is assigned resource only in TP $b$ whereas each user $u$ whose corresponding ratio $\frac{R_{u,1}}{R_{u,b}}$ is equal or greater than $\mu_{m,b}$ is assigned resource only by the Macro.
  The total  load imposed by users associated to TP $b$ on the Macro in this case is given by $\frac{N_b}{\lambda} - \frac{m-1}{\lambda}$.
 Then, note that irrespective of whether $\lambda \in \Sulc^A_b$ or $\lambda \in \Sulc^B_b$,
 upon increasing (decreasing) $\lambda$ the total load imposed on the Macro by users of TP $b$ decreases (increases), which establishes that the unique $\lambda$ that fully uses all the Macro resource can be determined using bi-section search to solve   relation in (\ref{eq:lrel}). \qed
\begin{figure}[!tbp]
  \centering
  \begin{minipage}[b]{0.45\textwidth}
    \includegraphics[width=\textwidth]{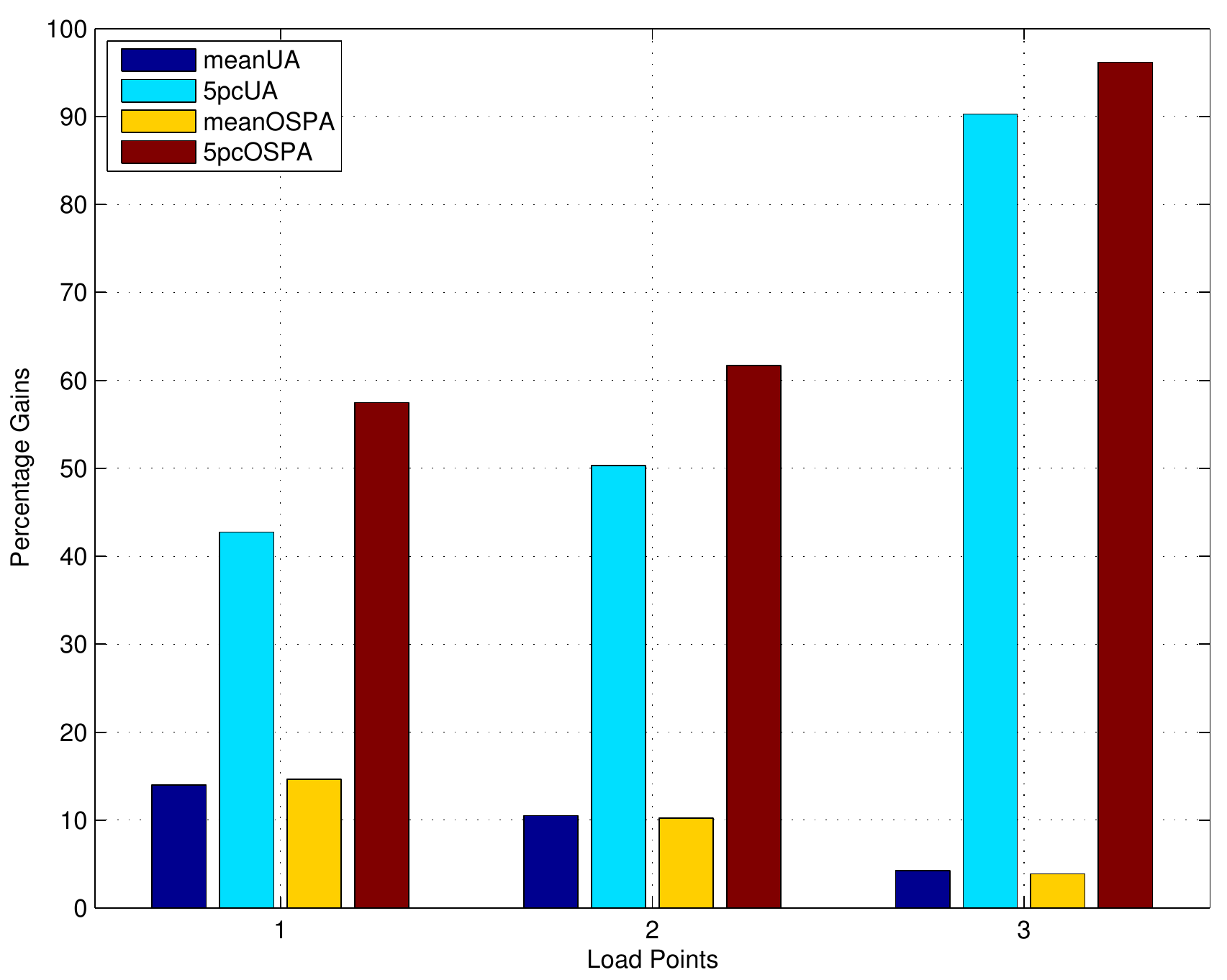}
    \caption{In-band scenario: pico and macro cells share same band.}\label{fig:DCIn}
  \end{minipage}
  \hfill
  \begin{minipage}[b]{0.45\textwidth}
    \includegraphics[width=\textwidth]{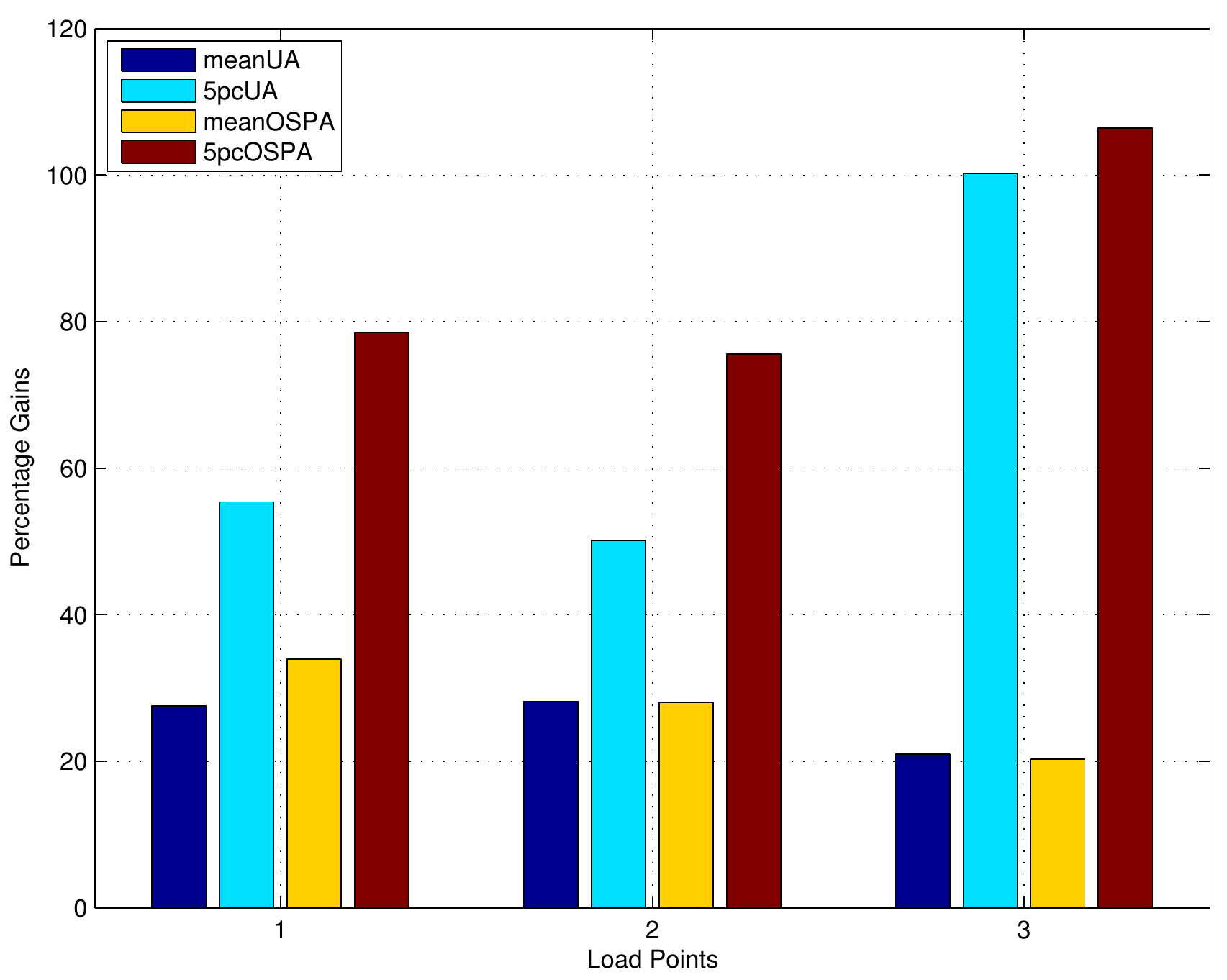}
    \caption{Out-band scenario: pico and macro cells share different bands.}\label{fig:DCOut}
  \end{minipage}\vspace{-.5cm}
\end{figure}


%

%
%


\end{document}